\documentclass[prb, aps, twocolumn, groupedaddress,showpacs]{revtex4}
\usepackage{eqnarray}
\usepackage[english]{babel} 
\usepackage[english]{layout}
\usepackage{amsmath,amsfonts,amssymb}
\usepackage{graphicx}
\usepackage{float}
\usepackage{color}
\usepackage{braket}
\usepackage{version}
\usepackage{array,multirow,makecell}

\begin{document}

\title{Quantum Valley Hall Effect and Perfect Valley Filter Based on Photonic Analogs of Transitional Metal Dichalcogenides}
\author{O. Bleu, D. D. Solnyshkov, G. Malpuech}
\affiliation{Institut Pascal, PHOTON-N2, University Clermont Auvergne, CNRS, 4 avenue Blaise Pascal, 63178 Aubi\`{e}re Cedex, France.} 

\begin{abstract}
We consider theoretically staggered honeycomb lattices for photons which can be viewed as photonic analogs of transitional metal dichalcogenides (TMD) monolayers. We propose a simple realization of a photonic Quantum Valley Hall effect (QVHE) at the interface between two TMD analogs with opposite staggering on each side. This results in the formation of valley-polarized propagating modes whose existence relies on the difference between the valley Chern numbers, which is a $\mathbb{Z}_2$ topological invariant.
We show that the magnitude of the photonic spin-orbit coupling based on the energy splitting between TE and TM photonic modes allows to control the number and propagation direction of these interface modes. Finally, we consider the interface between a staggered and a regular honeycomb lattice subject to a non-zero Zeeman field, therefore showing Quantum Anomalous Hall Effect (QAHE). In such a case, the topologically protected one-way modes of the QAHE become valley-polarized and the system behaves as a perfect valley filter.
\end{abstract}

\maketitle

\section{Introduction}

The band structure of a honeycomb lattice has been described in 1946 by Wallace within the tight-binding approximation \cite{Wallace1947}. The bands result from the coupling between the orbitals of the A and B atoms, and the Hamiltonian can be represented as an effective magnetic field acting on a pseudospin defined by the two sublattices (A and B). As widely advertised since the isolation of monolayer graphene, its dispersion contains two inequivalent Dirac points $K$ and $K'$, located at the corners of the Brillouin zone. The effective field cancels at these points, where A and B orbitals are uncoupled. Around these points, the effective field is monopolar at $K$ and Dresselhaus at $K'$: $H_{K,K'}=\hbar v_f(\tau_z q_x\sigma_x+q_y\sigma_y)$, where  $\tau_z=\pm1$ at $K$ and $K'$  respectively and $v_f=\frac{3aJ}{2\hbar}$ ($a$ is the lattice parameter and $J$ is the tunneling coefficient between the nearest neighbors). The two fields have opposite winding numbers ($\pm1$ respectively). A consequence is the $\pi$  and $-\pi$ Berry phase acquired by an electrons on a closed loop around the $K$ and $K'$ points respectively \cite{RevNiu2010}.

In a staggered honeycomb lattice, A and B sites have an energy difference \cite{Niu2007} $2\Delta$. In such a case, the effective field gains a Z component similar at both $K$ and $K'$ and the Hamitonian becomes $H_{K,K'}=\hbar v_f(\tau_z q_x\sigma_x+q_y\sigma_y)+\Delta\sigma_z$ . A gap opens between the bands, with the presence of a non-zero Berry curvature of opposite sign near $K$ and $K'$. As a result, the total Chern number of the bands is zero and they are topologically trivial. 
This Hamiltonian is realized by several real crystalline structures, such as the transitional metal dichalcogenides (TMD) monolayers \cite{Yao2012,Trushin2016}. A large part of the unique properties of this class of 2D materials is due to the local Berry curvature of their bands near the $K$ and $K'$ points and to the related orbital moment. For instance, it imposes specific optical selection rules. Only  circularly-polarized  $\sigma^+$ light can be absorbed by the $K$ valley transition, while only $\sigma^-$  light is absorbed by the $K'$ valley transition \cite{zeng2012valley,mak2012control}.  The valley dependent Berry curvature is also at the origin of the Valley Hall effect \cite{Niu2008,Yao2012,cao2012valley} recently observed experimentally in MoS$_2$ \cite{mak2014valley}. Electrons, accelerated by an electric field, experience a valley-dependent Berry curvature provoking a valley-dependent drift perpendicular to the acceleration direction, which leads to spatial separation of valley electrons.
 
The valley degree of freedom allows to define a valley pseudo-spin. Moreover, the Berry curvature is typically strongly concentrated around  $K$ and $K'$, which allows to define a valley Chern number $C_{K,K'}$ \cite{Niu2007} by analogy with the spin Chern number, which is linked to the $\mathbb{Z}_2$ topological invariant characterizing the quantum spin Hall effect (QSHE) \cite{kane2005z}. One might think that similarly with the QSHE effect, the presence of this non-zero topological invariant should lead to the existence of surface states (at the edges of the sample) with valley dependent chirality. This is not the case, because the Bloch wave functions associated with the valleys, are not the stationary eigenstates in the vacuum (outside the sample), contrary to the spin states, and are therefore mixed at such boundaries. However, at an interface (domain wall) between two TMD analogs inverted the one with respect to the other, zero-line modes are known to be present \cite{Semenoff2008}. These states can become valley-dependent and chiral with a proper choice of the lattice parameters \cite{Niu2009} leading to the so-called quantum valley Hall effect (QVHE).
  
While it is possible to invert a real TMD lattice simply by turning the 2D sheet upside down, the interfaces between two inverted lattices cannot be formed, because they would correspond to metal-metal or chalcogenide-chalcogenide chemical bonds. It seems therefore very difficult to obtain and study chiral states at such interfaces. Several theorical works have proposed to use bilayer graphene systems, where a tunable gap can be opened by applying a bias voltage \cite{McCann2006,GeimCastro2007} to organize these interface states \cite{Martin2008,li2011topological,zhang2013valley}. Chiral valley-polarized edge states are also predicted in these systems due to the coupling between the two layers with a potential difference \cite{Castro2008}. 
The signature of the existence of the interface modes has been reported experimentally in the bilayer system \cite{ju2015topological} and the study of valley polarized edge states or QVHE is nowadays a research field in several condensed matter systems \cite{Vaezi2013,Marino2015,ren2016topological,Yao2014}. 

TMD analogs can also be realized using cold atoms in optical lattices, or by coupling optical resonators. In the latter case, Berry curvature can be probed by resonant excitation of the Dirac point in momentum space and measurement of a drift of the wave packet, which, however, is predicted to be relatively small \cite{Carusotto2014}. Recently, photonic quantum valley Hall effect in photonic crystal slabs \cite{barik2016two,Shvets2015,ma2016all,chen2016valley} have been theoretically described. In these works  chiral Valley polarized interface states are organized, but are not protected against disorder induced inter-valley scattering. 

  
  In this work, we first propose to implement Quantum Valley Hall effect in a photonic system using a honeycomb lattice of coupled micropillars \cite{Jacqmin2014}, with a tunable energy detuning between A and B pillars. The gap size can be entirely controlled by tuning the pillar diameter. In the second section, we consider the impact of the splitting between TE and TM optical modes, which is specific of photonic systems which was not considered before. We show that the magnitude of the splitting allows to tune the direction of propagation and number of interface modes. In the last section we propose an original scheme to build a perfect valley filter by using a domain wall between QAH and QVH phases using mixed light-matter exciton-polariton quasi-particles. 
The QAH phase for polaritons has been predicted to occur in different kind of polariton lattices under Zeeman field and is of current interest \cite{polaritonZ2015,KarzigPRX2015,LiewPRB2015,gulevich2016kagome}. Obviously, the model used in this work is based on the honeycomb lattice.  The dispersion of the interface modes as well as their real space propagation can be directly measured by optical spectroscopy techniques. The valley structure of a wavepacket can be analyzed and controlled as well.

\begin{figure}[tbp]
 \begin{center}
 \includegraphics[scale=0.42]{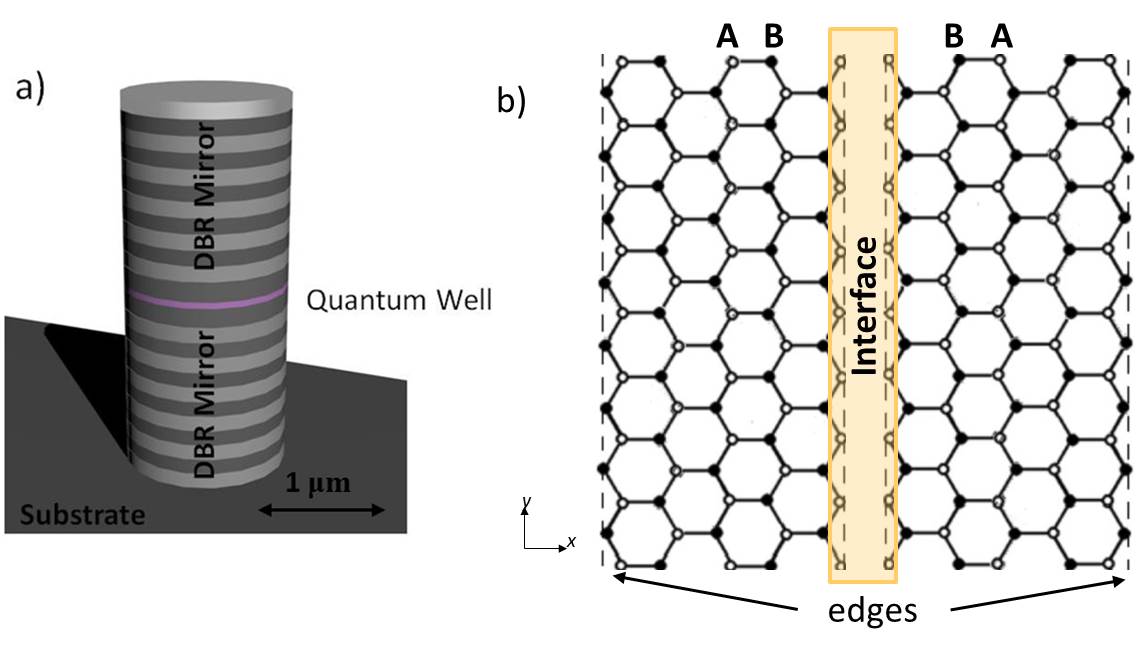}
 \caption{(a) Cavity micropillar (artificial atom) scheme. (b) Zigzag interface between two TMD analog lattices with opposite A-B organization giving rises to zero lines modes and quantum valley Hall effect for good parameters.}
  \end{center}
 \end{figure}

\section{Photonic Quantum Valley Hall Effect}

\begin{figure}[tbp]
 \begin{center}
 \includegraphics[scale=0.57]{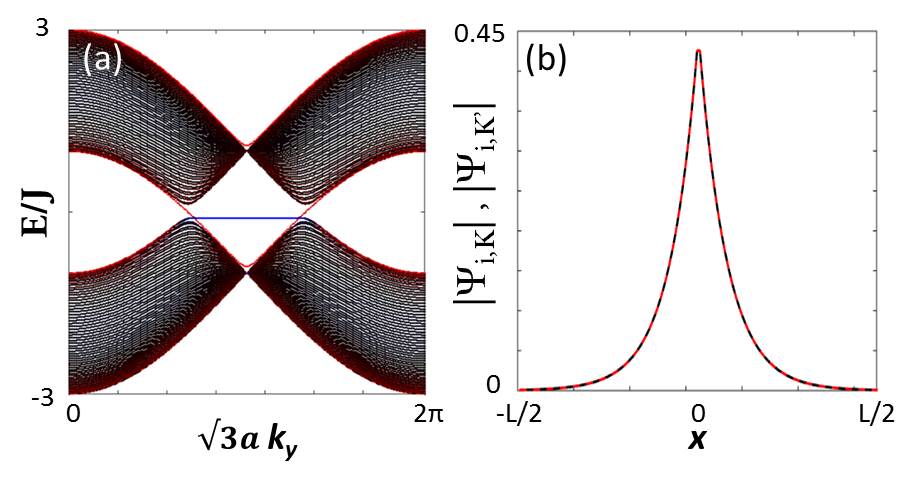}
 \caption{(Color online) (a) Ribbon dispersion (Colors represent localization on the interface (red) and on the edges of the structure (blue)). (b) Corresponding in-gap interface wavefunction absolute values projection on the transverse (x) direction for $K$ (solid-black) and $K'$ (dashed-red) valleys. (Staggered potential: $\Delta_{AB}(r)=-\Delta_{AB}(l)=0.1 J$)}
  \end{center}
 \end{figure}
 
An interesting aspect of using lattices of coupled photonic micropillars, like the one shown in Fig.~1(a), is that the tight-binding description is a very good approximation. The insertion of the quantum well in the optical cavity allows to achieve strong light matter coupling and to obtain exciton-polariton (polariton) eigenstates \cite{kavokin2011microcavities}. Thanks to their excitonic part, polaritons shows a sizeable Zeeman splitting between the two circular-polarized components under applied magnetic field in the growth direction or in the high density regime via spin-anisotropic interactions \cite{shelykh2009polariton}. 

However, we will at first neglect the polarization degree of freedom. In this case, the tight-binding Hamiltonian describing the photonic states in the lattice of pillar cavities can be written as:
\begin{equation}
H_k=\begin{pmatrix}
\Delta_{AB} && -Jf_k\\
-Jf_k^* && -\Delta_{AB}
\end{pmatrix}, ~~ f_k=\sum_{j=1}^3\exp{(-i\textbf{kd}_{\phi_j})} 
\end{equation}
where $2\Delta_{AB}$ is the energy difference between the ground states of A and B pillars and $J$ is the nearest neighbours tunnelling coefficient. A non-zero $\Delta_{AB}$ leads to the opening of a band gap and induces opposite Berry curvatures around $K$ and $K'$ points, which for the "valence band" reads: 
\begin{equation}
B(q)=\tau_z\frac{9a^2J^2\Delta_{AB}}{(4\Delta_{AB}^2+9a^2J^2q^2)^{3/2}}.
\end{equation}
Here, $\tau_z=\pm1$ is the valley index. 
If $\Delta_{AB}$ is small, the Berry curvature is strongly localized around $K$ points allowing to define the valley Chern numbers as  $C_{K,K'}=\pm0.5$. The total Chern number of the band $C=C_{K}+C_{K'}$ is zero. The flexibility of photonic systems allows to realize staggered graphene lattices with any sign of $\Delta_{AB}$. The scheme proposed here is therefore to build an interface between two TMD analogs with opposite energy bandgaps ($\Delta_{AB}<0$ for $x<0$ and $\Delta_{AB}>0$ for $x>0$) as first proposed theoretically in \cite{Semenoff2008}.  We stress that this kind of interface is not experimentally realistic for real electronic TMD monolayers. On the contrary, for the photonic analog that we propose and consider here, the gap sign is controlled by the sizes of the A and B micropillars. Such lattices and interfaces are very well within the current experimental possibilities \cite{Jacqmin2014,milicevic2015edge}.

We compute the dispersion of a ribbon with such interface (which can be considered as a domain wall) along the zigzag direction (see scheme Fig.~1(b)) using the tight-binding approach. The ribbon of size $L$ is constructed with 128 coupled zigzag infinite chains, the interface is located in the middle of the ribbon. The dispersion of the full ribbon is plotted in Fig.~2 (a), where the interface states are shown with red and and the real boundary modes are shown with blue (both for left and right edges, which are degenerate). The latter are non-dispersive, connecting the extrema at the top of the valence band. On the other hand, the interface modes (red) are dispersive. They connect the valence and the conduction band in each valley, with group velocities opposite in the two valleys. Hence, this "zero-line mode", as commonly called in the literature \cite{Semenoff2008,Martin2008,Feng2013}, is associated with a QVH current on the interface.

The presence of the QVH current can be understood with a topological argument based on the difference of the valley Chern numbers across the interface $N_{K,K'}=C_{K,K'}(l)-C_{K,K'}(r)=\pm 1$ (where $l$ and $r$ stand for the left and right domains) \cite{Martin2008,Li2010}. One should note that the valley Chern number is different from the Chern number of a full band. The topological charge $N$ is used to characterize zero-line modes between two 2D Dirac vacua with inverted mass gaps \cite{volovik2003universe}. 
In honeycomb lattices, there are two inequivalent valleys with Dirac dispersions, leading to two zero-line modes (one for each valley) at the interface.

 The overlap of the wavefunctions of the interface states (corresponding to the wavevectors $K$ and $K'$) in real space in the direction perpendicular to the interface is perfect, as shown in Fig.~2(b).  This implies that these states are not topologically protected against scattering on a defect between the $K$ and $K'$ valleys. In electronic systems, the practical argument which is put forward is that in sufficiently clean systems this kind of backscattering tends to be negligible because of the relatively large distance between the two valleys in momentum space. This arguments fails in photonics, since the valleys are, on the opposite, relatively close in reciprocal space. Only some specific defect shapes, suppressing the scattering in particular directions and thus suppressing the coupling between the valleys, would keep the valley current unperturbed. On the other hand, a random disorder, such as the variation of the pillar diameter in micropillar lattices, or the variation of the hole size in photonic crystal slabs certainly induces an important scattering between the valleys. In the last section of the manuscript we demonstrate the absence of topological protection for the valley current in the QVHE by simulating the scattering on a localized defect.

\section{Quantum Valley Hall Effect with photonic Spin-Orbit Coupling}

We now include the light polarization degree of freedom and the so-called TE-TM spin-orbit coupling \cite{CRAS2016} (present in any photonic system) in our model. The Hamiltonian becomes a $4 \times 4$ matrix \cite{Anton2015} written here in the $\left( \Psi_A^+,\Psi_A^-,\Psi_B^+,\Psi_B^-\right)$ basis.
\begin{eqnarray}
H_{qvh}=\begin{pmatrix}
\Delta_{AB} \mathbb{I}  &F_k \\
F_k^\dagger& -\Delta_{AB}\mathbb{I} 
\end{pmatrix} 
,\quad F_k=-\begin{pmatrix}
f_kJ&f_k^+\delta J \\
f_k^- \delta J&f_kJ 
\end{pmatrix}
\end{eqnarray}
where $\delta J$ is the spin orbit coupling strength and $f_k^{\pm}$ coefficients are defined by: $f_k^{\pm}=\sum_{j=1}^3 \exp{(-i[\textbf{kd}_{\phi_j}\mp 2\phi_j])}$.
The definition of valley Chern number when spin-orbit coupling is present is not strict because the SOC brings an additional contribution to the Berry curvature of each band \cite{bleu2016effective} which can lead to non-integer $C_K$. However, if  the valleys are still energetically defined (which means that they are clearly visible in the dispersion), and the two lattices on each side of the interface are perfectly inverted, the difference between valley Chern numbers $N_{K,K'}$ is a well defined integer. 

\subsection{Valley topological charges: analytical results}
To demonstrate that the difference between the valley Chern numbers remains integer, we  compute analytically the topological charges of the valleys in a staggered honeycomb lattice with TE-TM SOC. For this purpose we derive an effective low-energy Hamiltonian to describe the low energy bands (closest to the central gap) in a given valley.
Indeed, a $2\times 2$ Hamiltonian with particle-hole symmetry can be written as a superposition of Pauli matrices: $H_{eff}=\mathbf{\Omega}.\mathbf{\sigma}$ where $\mathbf{\Omega}$ is an effective magnetic field.
Using this notation, the Berry curvature can be written as: 
\begin{equation}
\mathbf{B}=\frac{1}{2|\mathbf{\Omega}|^3}\mathbf{\Omega}.(\partial_{q_x}\mathbf{\Omega} \times\partial_{q_y}\mathbf{\Omega})
\label{windingberry}
\end{equation}
whose integral $\frac{1}{2\pi}\int\mathbf{B}\mathrm{d} \mathbf{q}$ correspond to the winding number of the pseudospin on the Bloch sphere which is equivalent to the Chern topological invariant for $2\times2$ Hamiltonians.

In order to use the above picture, we need to reduce the number of states in the basis of our system. First, we linearize the $4\times 4$ Hamiltonian defined above around the $K$ point ($0,\frac{4\pi}{3\sqrt{3}a}$). Using the new coordinate $\mathbf{q}=(\mathbf{k}-\mathbf{K})$, we rewrite it in the basis $\left(\Psi_A^+,\Psi_B^-,\Psi_A^-,\Psi_B^+\right)$ where $A/B$ and $\pm$ are the sublattice and spin indices respectively.
\begin{widetext}
\begin{equation}
H_{qvh,K}=\begin{pmatrix}
\Delta_{AB} && -\frac{3a\delta J}{2} (q_y-iq_x) && 0&& \frac{3aJ}{2} (q_y+iq_x)\\ -\frac{3a\delta J}{2} (q_y+iq_x) &&-\Delta_{AB}&&\frac{3aJ}{2} (q_y-iq_x)&&0\\ 0&&\frac{3aJ}{2} (q_y+iq_x)&&\Delta_{AB}&&-3\delta J\\\frac{3aJ}{2} (q_y-iq_x) && 0&&-3\delta J&& -\Delta_{AB}
\end{pmatrix}=\begin{pmatrix}
H_1&&T\\T&&H_2
\end{pmatrix}
\end{equation}
\end{widetext}
We are interested in the two  branches closest to the gap (the inner branches of the dispersion). The perturbation theory is applicable if the high energy branches are sufficiently far away, which means $\delta J>> \Delta_{AB}$. Using the perturbation theory, one can derive an effective $2\times 2$ Hamiltonian: $H_{eff}=H_1-TH_2^{-1}T$.
In the following, we are interested in two different limits: when $\delta J << J$ and when $\delta J \sim J$.

In the first limit, one can neglect the term $\sim\delta J q$ (corresponding to neglect the trigonal warping \citep{Anton2015}). The effective Hamiltonian can then be written as:
\begin{widetext}
\begin{equation}
H_{K,eff}^{(1)}=\begin{pmatrix}
\Delta_{AB} &&0 \\ 0&& -\Delta_{AB}
\end{pmatrix}+ \frac{1}{\Delta_{AB}^2+9\delta J^2}\begin{pmatrix}
\frac{9a^2J^2\Delta_{AB}}{4}q^2 &&\frac{27a^2J^2\delta J}{4}(q_y+iq_x)^2 \\ \frac{27a^2J^2\delta J}{4}(q_y-iq_x)^2&& -\frac{9a^2J^2\Delta_{AB}}{4}q^2 
\end{pmatrix}
\end{equation}
\end{widetext}
Due to the absence of the terms $\sim\delta J q$, when $\Delta_{AB}=0$ the energies are two inverted parabola degenerate at $\mathbf{q}=0$. A similar development around $K'$ ($0,-\frac{4\pi}{3\sqrt{3}a}$) allows to obtain $H_{K',eff}$.
Using Eq. \eqref{windingberry}, we compute the Berry curvature in this limit: 
\begin{equation}
\mathbf{B}^{(1)}=\frac{\tau_z144 a^4 \Delta_{AB} J^4\delta J^4 q^2}{ (
  9 a^4 \delta J^2 J^4 q^4 + \Delta_{AB}^2 (4 \delta J^2 + a^2 J^2 q^2)^2)^{3/2}}\mathbf{e_z}
\end{equation}
where $\tau_z=\pm1$ for $K$/$K'$ valleys. 
The corresponding valley Chern number is:
\begin{eqnarray}
C_{K,K'}^{(1)}=\frac{1}{2\pi}\int\mathbf{B}\mathrm{d} \mathbf{q}=\tau_z\mathrm{sign}(\Delta_{AB}) (1-  \frac{\tau_z\Delta_{AB}}{\sqrt{\Delta_{AB}^2+9\delta J^2}}) 
\label{vcn}
\end{eqnarray}
The valley Chern numbers tend to $C_{K,K'}^{(1)}=\tau_z\mathrm{sign}(\Delta_{AB})$ if the high energy bands are far away from the low energy ones ($\Delta_{AB}<<\delta J$). As a consequence, the topological invariant characterizing the domain wall between two inverted TMD is: 
\begin{equation}
 N_{K,K'}^{(1)}=C_{K,K'}(l)-C_{K,K'}(r)= \pm2 \mathrm{sign}(\Delta_{AB}) 
\end{equation} 
 which means that there are two co-propagating interface states in a given valley.
One can see from Eq. \eqref{vcn}, that if the high energy bands are too close, the valley Chern number of a given low energy band is not an integer. This  means that the topological charge is shared between the high and low energy bands. In this case, perturbation theory becomes inapplicable and one needs to compute the valley Chern numbers numerically.

In the second limit, where $\delta J \sim J$,  the linear terms ($\delta J q$) have to be conserved and we can neglect the quadratic terms $q^2$. The resulting effective $2\times 2$ Hamiltonian around $K$ and $K'$ can be written in the ($\Psi_A^+,\Psi_B^-$) and ($\Psi_A^-,\Psi_B^+$) basis:

\begin{eqnarray}
H_{K,eff}^{(2)}=\begin{pmatrix}
\Delta_{AB} &&\frac{3a\delta J}{2}(q_y-iq_x) \\ \frac{3a\delta J}{2}(q_y+iq_x) && -\Delta_{AB}
\end{pmatrix}\\ H_{K',eff}^{(2)}=\begin{pmatrix}
\Delta_{AB} &&-\frac{3a\delta J}{2}(q_y+iq_x) \\ -\frac{3a\delta J}{2}(q_y-iq_x) && -\Delta_{AB}
\end{pmatrix}
\end{eqnarray}
By looking on these two effective Hamiltonians, we can directly see that the signs of the diagonal elements are the same, whereas the windings of the effective in-plane fields are opposite, exactly as in staggered honeycomb lattices without SOC.
Hence, the Berry curvatures are opposite in each valleys. They can be calculated as: 
\begin{equation}
\mathbf{B}^{(2)}=-\frac{\tau_z 9 a^2 \Delta_{AB} \delta J^2}{(4 \Delta_{AB}^2 + 9 a^2 \delta J^2 q^2)^{3/2}}\mathbf{e_z}
\end{equation}
where $\tau_z=\pm1$ for $K$/$K'$ valleys. The corresponding valley Chern numbers are $C_{K,K'}^{(2)}=- \frac{\tau_z}{2}\mathrm{sign}(\Delta_{AB})$.
The domain wall invariant between two inverted QVH phases is then $N_{K,K'}^{(2)}=\mp\mathrm{sign}(\Delta_{AB})$ which implies the presence of one interface mode in each valley. The group velocities of these modes in the two valleys are of course opposite.

This low energy study has allowed us to determine analytically the valley topological charge in these two limits and hence to compute the number of valley-polarized interface states. One should note that the topological charge of a given valley changes sign between the two limits. The group velocities of the interface states are inverted as well. 
In the appendix, we show the effective Hamiltonian derived in the general case between the two limits considered here.

\subsection{Spectrum and discussion}

\begin{figure}[tbp]
 \begin{center}
 \includegraphics[scale=0.52]{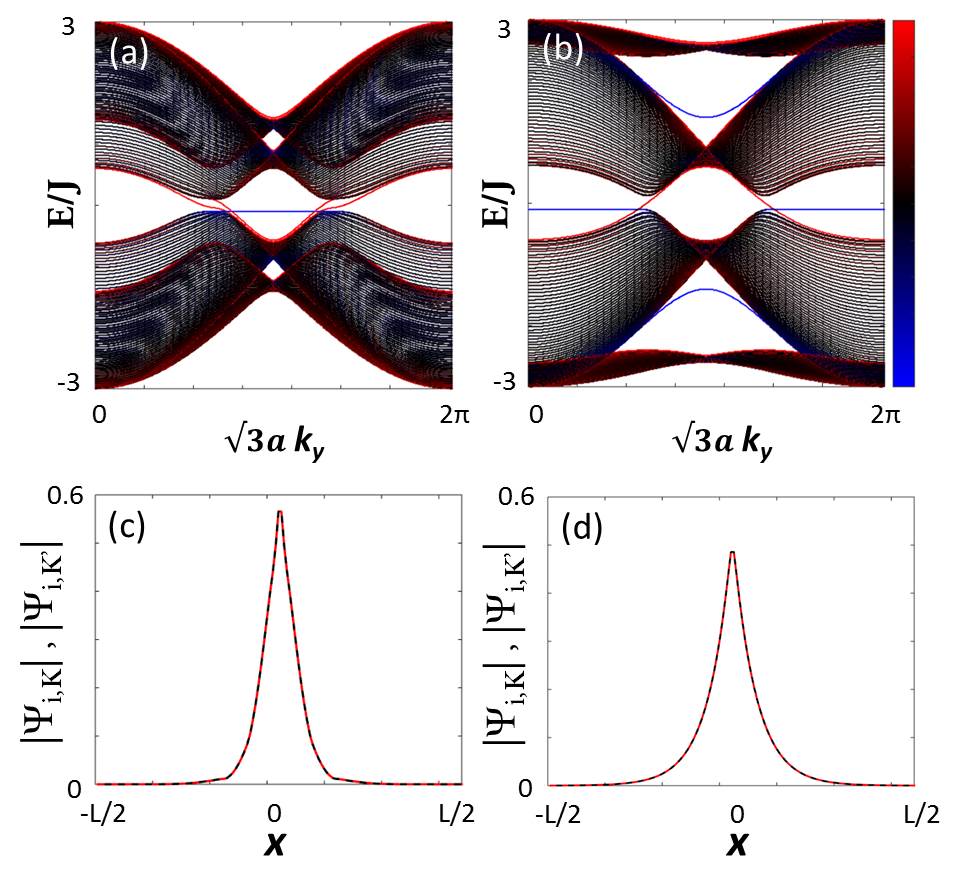}
 \caption{(Color online) (a,b) Ribbon dispersions with TE-TM SOC: (a) $\delta J=0.2J$ and (b) $\delta J=0.8J$. Colors represent localization on the interface (red) and on the edges of the structure (blue). (c,d) Corresponding absolute values of the interface wavefunctions at $K$ (solid-black) and $K'$ (dashed-red) points. (Staggered potential: $\Delta_{AB}(r)=-\Delta_{AB}(l)=0.1 J$)}
  \end{center}
 \end{figure}

In the previous subsection, we have derived an effective low energy two-band Hamiltonian in order to compute analytically the valley topological charges. We have shown that it is possible to distinguish between two configurations. In the weak SOC limit, $N_{K,K'}=\pm2 \mathrm{sign}(\Delta_{AB})$, whereas when $\delta J \sim J$, $N_{K,K'}=\mp \mathrm{sign}(\Delta_{AB})$.
Here, we compute the corresponding tight-binding dispersions of the ribbon which are shown in Fig.~3(a) and (b) respectively. In Fig.~3(a), there are two interface states in each valley, that are visible within the bandgap. The ribbon dispersion in the second limit is shown in Fig.~3(b), this time there is only one interface state in each valley, as predicted from the topological invariant. 
The corresponding spatial overlap of the interface wavefunctions at $K$ and $-K$ is shown in Fig.~3(c,d). As in the case without SOC, the backscattering from one valley to the other is not forbidden. However, it should be noted that the presence of TE-TM SOC induces a polarization mismatch between the two valleys (not visible on these figures, showing only the absolute value of the wavefunction), which reduces the exact overlap of the two wavefunctions.

\section{Perfect optical valley filter}

Exciton-polaritons, mixed exciton-photon quasi-particles appearing in microcavities in the strong coupling regime, are sensitive to an external magnetic field due to their excitonic part. Optical quantum anomalous Hall (QAH) effect has been predicted to occur in honeycomb lattice of polariton micropillars with TE-TM SOC under a Zeeman field \cite{polaritonZ2015,bleu2017photonic}.  In this section, we propose to organize an interface between this QAH phase and the QVH phase presented above, in order to obtain a perfect valley filter as sketched in Fig.~4(a), where the colored arrows represent the propagating edge states on the boundaries of the QAH phase.
A similar scheme has been recently theoretically proposed for electron in graphene with Rashba SOC \cite{Pan2015}. However, we stress that the QAH effect has not been observed yet in graphene and that the creation of such interface in electronic systems seems quite challenging experimentally.

\subsection{QAH phase: analytical results}

The tight-binding Hamiltonian describing the photonic QAH phase is the following \cite{polaritonZ2015}:
\begin{eqnarray}
H_{qah}=\begin{pmatrix}
\Delta_{z} \sigma_z  &F_k \\
F_k^\dagger & \Delta_{z}\sigma_z 
\end{pmatrix} 
\end{eqnarray}
where $\Delta_z$ is the Zeeman splitting. 
We propose to create a topological domain wall between QAH and QVH phases. 
In this subsection we derive an effective low energy theory for the QAH phase. The Hamiltonian around the $K$ point can be written in the $\left(\Psi_A^+,\Psi_B^-,\Psi_A^-,\Psi_B^+\right)$ basis:
\begin{widetext}
\begin{equation}
H_{qah,K}=\begin{pmatrix}
\Delta_{z} && -\frac{3a\delta J}{2} (q_y-iq_x) && 0&& \frac{3aJ}{2} (q_y+iq_x)\\ -\frac{3a\delta J}{2} (q_y+iq_x) &&-\Delta_{z}&&\frac{3aJ}{2} (q_y-iq_x)&&0\\ 0&&\frac{3aJ}{2} (q_y+iq_x)&&-\Delta_{z}&&-3\delta J\\\frac{3aJ}{2} (q_y-iq_x) && 0&&-3\delta J&& \Delta_{z}
\end{pmatrix}=\begin{pmatrix}
H_1&&T\\T&&H_2
\end{pmatrix}
\end{equation}
\end{widetext}
The difference between this linearized Hamiltonian and the one of QVH phase is that the signs of the diagonal elements of $H_2$ are inverted.
In the configuration, where a two-band effective theory can be applied ($\delta J>>\Delta_z$), we study the same limits as in the previous section: $\delta J << J$ and $\delta J \sim J$. In the first limit, one has: 
\begin{widetext}
\begin{equation}
H_{eff}^{(1)}=\begin{pmatrix}
\Delta_{z} &&0 \\ 0&& -\Delta_{z}
\end{pmatrix}+ \frac{1}{\Delta_{z}^2+9\delta J^2}\begin{pmatrix}
-\frac{9a^2J^2\Delta_{z}}{4}q^2 &&\frac{27a^2J^2\delta J}{4}(q_y+iq_x)^2 \\ \frac{27a^2J^2\delta J}{4}(q_y-iq_x)^2&& \frac{9a^2J^2\Delta_{z}}{4}q^2 
\end{pmatrix}
\end{equation}
\end{widetext}
which implies the following Berry curvature when $\delta J>>\Delta_z$: 

\begin{equation}
\mathbf{B}^{(1)}=\frac{144 a^4 \Delta_{z} J^4\delta J^4 q^2}{ (
  9 a^4 \delta J^2 J^4 q^4 + \Delta_{z}^2 (a^2 J^2 q^2-4 \delta J^2)^2)^{3/2}}\mathbf{e_z}
\end{equation}
A similar development at $K'$ point giving the same result, one can deduce the two valleys Chern numbers:

\begin{eqnarray}
C_{K,K'}^{(1)}=\mathrm{sign}(\Delta_{z}) (1+ \frac{\Delta_{z}}{\sqrt{\Delta_{z}^2+9\delta J^2}}) \approx \mathrm{sign}(\Delta_{z})
\end{eqnarray}
When $\delta J \sim J$, the low energy $2\times 2$ Hamiltonians around $K$ and $K'$ can be written in ($\Psi_A^+,\Psi_B^-$) and ($\Psi_B^+,\Psi_A^-$) effective basis respectively as:
\begin{eqnarray}
H_{K,eff}^{(2)}=\begin{pmatrix}
\Delta_{z} &&\frac{3a\delta J}{2}(q_y-iq_x) \\ \frac{3a\delta J}{2}(q_y+iq_x) && -\Delta_{z}
\end{pmatrix}\\ \quad~~H_{K',eff}^{(2)}=\begin{pmatrix}
\Delta_{z} &&-\frac{3a\delta J}{2}(q_y-iq_x) \\ -\frac{3a\delta J}{2}(q_y+iq_x) && -\Delta_{z}
\end{pmatrix}
\end{eqnarray}

We can see that both the Zeeman field and the winding of the in-plane field have the same sign which results in equal Berry curvatures:
\begin{equation}
\mathbf{B}^{(2)}=-\frac{9 a^2 \Delta_{z} \delta J^2}{(4 \Delta_{z}^2 + 9 a^2 \delta J^2 q^2)^{3/2}}\mathbf{e_z}
\end{equation}
leading to valley topological charges $C_{K,K'}^{(2)}=-\frac{1}{2}\mathrm{sign}(\Delta_{z}) $. 
These analytical results are useful to understand the behavior of the topologically protected edge states of a quantum anomalous Hall phase at the interface with a QVH phase, where the two valleys are not equivalent.
Using the domain wall topological invariant $N_{K,K'}$ allows to predict the number of \textit{topologically protected} interface states in a given valley. Of course, this topological argument is valid only when the valleys are energetically well defined.

\subsection{Spectrum and discussion}

\begin{figure}[bp]
 \begin{center}
 \includegraphics[scale=0.48]{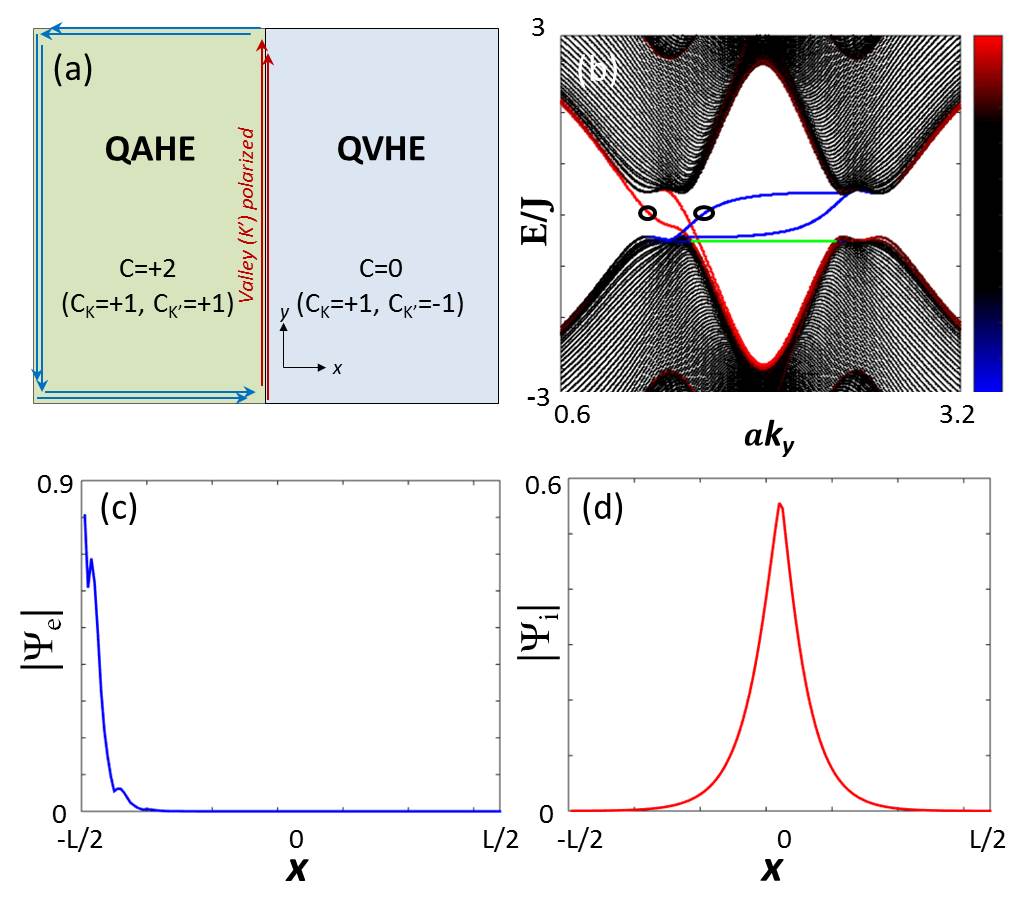}
\caption{(Color online) (a) Scheme of the structure, (b) Ribbon dispersion (Colors represent localization on the interface (red) and on the left edge of the structure (blue)). (c,d) Wavefunction projections on the transverse (x) direction for the states highlighted by the circles in (b): (c) edge (d) interface.  (Parameters: $\Delta_{z}(l)=\Delta_{AB}(r)=0.1 J$, $\delta J=0.2 J$, $\Delta_{z}(r)=\Delta_{AB}(l)=0$)} 
\end{center}
 \end{figure}
 
As above, we use a tight-binding approach in order to find the edge states. The two domains are described by $H_{qah}$ and $H_{qvh}$, respectively. We consider a realistic configuration for a lattice of micropillars where the TE-TM SOC is relatively weak ($\delta J < J/2$). 

In the left phase, the total Chern number is equal to $C(l)=2\times\mathrm{sign}(\Delta_z)$ whereas in the right one $C(r)=0$, as explained above. From the bulk-boundary correspondence, the interface supports two modes propagating in the same direction, given by the sign of $\Delta_z$.
When the valleys are energetically well defined, one can compute the valley Chern numbers in each side analytically (previous subsection) or numerically. In the weak $\delta J$ limit, $C_{K}(l)=C_{K'}(l)=\mathrm{sign}(\Delta_z)$ in the left domain, whereas $C_{K}(r)=-C_{K'}(r)=\mathrm{sign}(\Delta_{AB})$ in the right domain.
The resulting domain wall invariants for each valley are:
\begin{equation}
N_{K,K'}=\mathrm{sign}(\Delta_z)\mp\mathrm{sign}(\Delta_{AB})
\end{equation}
We stress here that even if the valley Chern numbers are not well defined integers for all range of parameters, their difference $N_{K,K'}$ defined above is always an integer.
Therefore, the domain wall invariant always cancels for one valley, and is non-zero for the other valley. The two one-way modes are therefore always valley polarized.

To confirm this analysis based on topological argument, we compute the dispersion of a semi-infinite ribbon as we did in the previous sections. The results are shown in Fig.~4. In panel (b), we can see that the interface states (red) are localized in one of the two valleys in momentum space. The non-dispersive edge states of the QVH phase are shown in green. The one-way dispersive edge states of the QAH phase (blue) cross the gap and connect the two valleys. Contrary to the interface between two QVH phases, there is no interface state crossing the gap in the second valley. This means that the valley polarized interface states are really topologically protected against backscattering in the presence of defects. As sketched in Fig.~4(a), the topologically protected edge states are purely valley polarized due to the domain wall topological invariant $N$. To clarify the difference between this scheme and the interface between two QVH phases described in the previous section, we plot in Fig.~4(c,d) the \emph{edge} and \emph{interface} state wavefunction (absolute values) projections in the $x$ direction. We can see that the blue state is clearly localized on the left edge of the structure and only the red ones are localized on the interface. This visual argument allows to understand the topological protection due to the absence of spatial overlap between the two wavefunctions contrary, to the case presented in the previous sections (Fig. 3(c,d)).

\section{Numerical simulations}

In this part, we present the full numerical simulations using the Schr\"odinger equation that we have performed in order to confirm the different results obtained in the tight-binding approach.
We describe the evolution of polaritons in the honeycomb lattice potential by solving the spinor Schr\"odinger equation for polaritons (in the parabolic approximation close to the bottom of the LPB).
\begin{eqnarray}\label{schro}
& i\hbar \frac{{\partial \psi _ \pm  }}
{{\partial t}}  =  - \frac{{\hbar ^2 }}
{{2m}}\Delta \psi _ \pm  - \frac{{i\hbar }}
{{2\tau }}\psi _ \pm  \pm\Delta_z\psi_\pm\\
& + \beta {\left( {\frac{\partial }{{\partial x}} \mp i\frac{\partial }{{\partial y}}} \right)^2}{\psi _ \mp } +U\psi_\pm + \hat{P} \nonumber 
\end{eqnarray}
where ${\psi_+(\mathbf{r},t), \psi_-(\mathbf{r},t)}$ are the two circular components, $m=5\times10^{-5}m_{el}$ is the polariton mass, $\tau=30$ ps the lifetime, $\beta$ is the TE-TM coupling constant (corresponding to a 5\% difference in the longitudinal and transverse masses). $\Delta_z$ is the magnetic field (applied only in the "Z" (QAH) region), $U$ is the lattice potential (radius of the pillars $r=1.5\mu m$, lattice parameter $a=2.5\mu m$), and  $\hat{P}$ is the pump operator.
We have generated 2 lattice potentials: one for staggered honeycomb lattice and one for unperturbed honeycomb lattice. In the staggered lattice, the deviation of the pillar radius from the average value was chosen to be 15\%.

Depending on the configuration (QVH/QVH phases or QVH/QAH phases), we either excite a large spot with a well-defined superposition of wavevectors corresponding to a single valley of the TMD lattice (QVH phase), or a single pillar of the Z insulator (QAH phase), but always with the frequency within the bulk gap.

\begin{figure}[tbp]
 \includegraphics[scale=0.57]{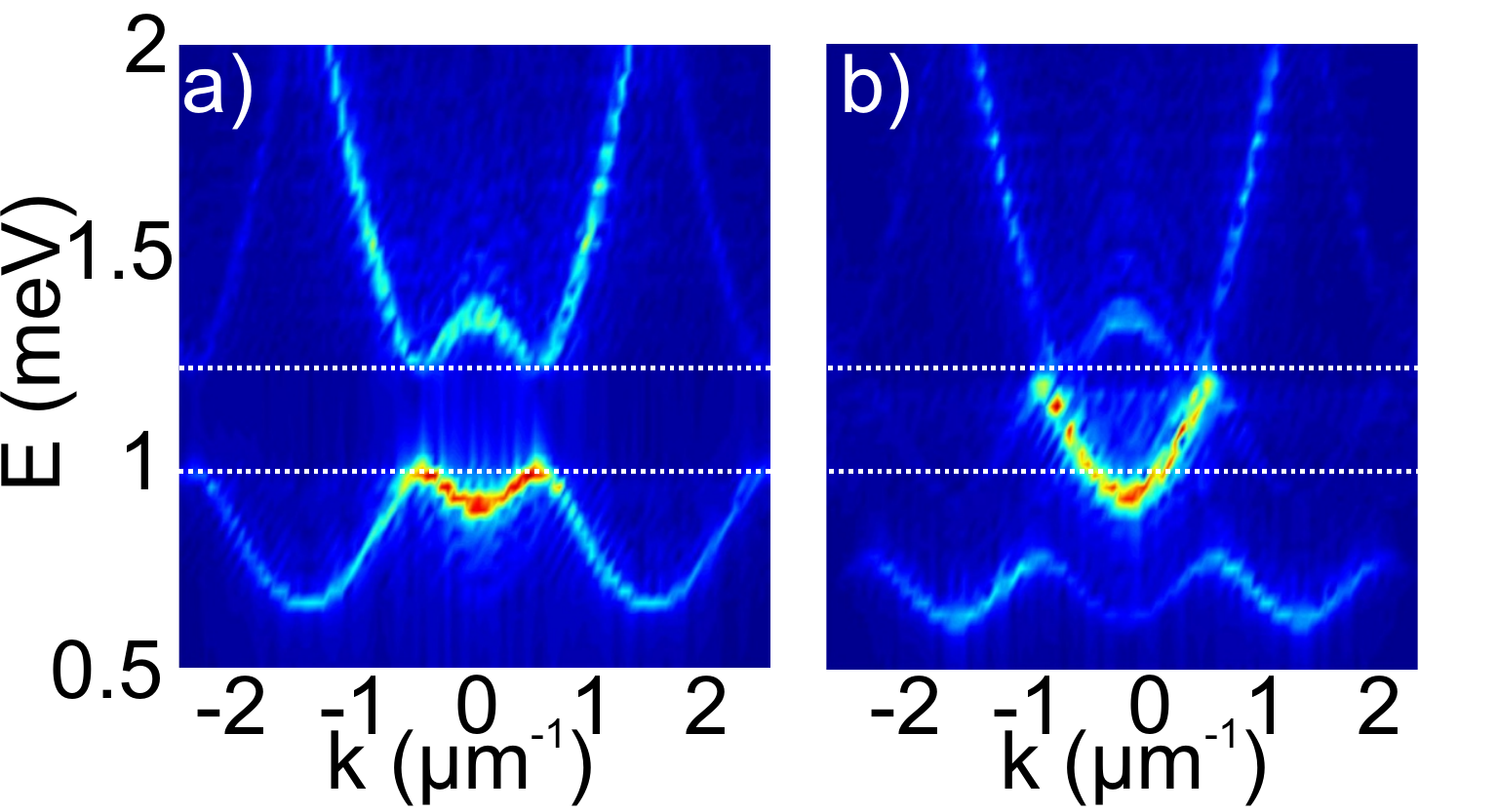}
\caption{(Color online) Dispersion of the a) bulk TMD and b) the TMD/TMD interface states in its gap. The gap is marked by a dashed white line.} 
\label{figdisp}
 \end{figure}

Figure \ref{figdisp}(a) shows the dispersion of the bulk TMD analog lattice. The image is a cut of a 2D dispersion in the $KMK'$ direction. Such representation allows to better visualize the edge of the Brillouin zone and the gap (shown by dashed white line). Panel (b) shows the dispersion of the states localized on the interface between two mutually inverted TMD analog lattices. Two large pillars are joined at the interface, which creates a potential trap even for the bulk states, whose dispersion is also visible below the bulk band. However, most interesting are the states in the central gap, which clearly exhibit opposite group velocities for the $K$ and $K'$ valleys. To excite these interface states, the laser spot was localized at the interface. This also suppresses the excitation of the bulk, whose dispersion therefore does not appear on panel (b).
 
While the TE-TM splitting was neglected for the calculation of the dispersion in Fig. \ref{figdisp} in order to obtain a clear image with a higher resolution, we have then included the TE-TM splitting in all numerical calculations presented below, both for QVH/QVH and QVH/QAH interfaces. All spatial images shown below are snapshots taken from the corresponding video files available as a Supplemental material for the manuscript \cite{suppl}. These spatial images show the total intensity of emission, given by $|\psi_{+}|^2+|\psi_-|^2$, as a function of spatial coordinates $x$ and $y$.

\begin{figure}[tbp]
 \includegraphics[scale=0.33]{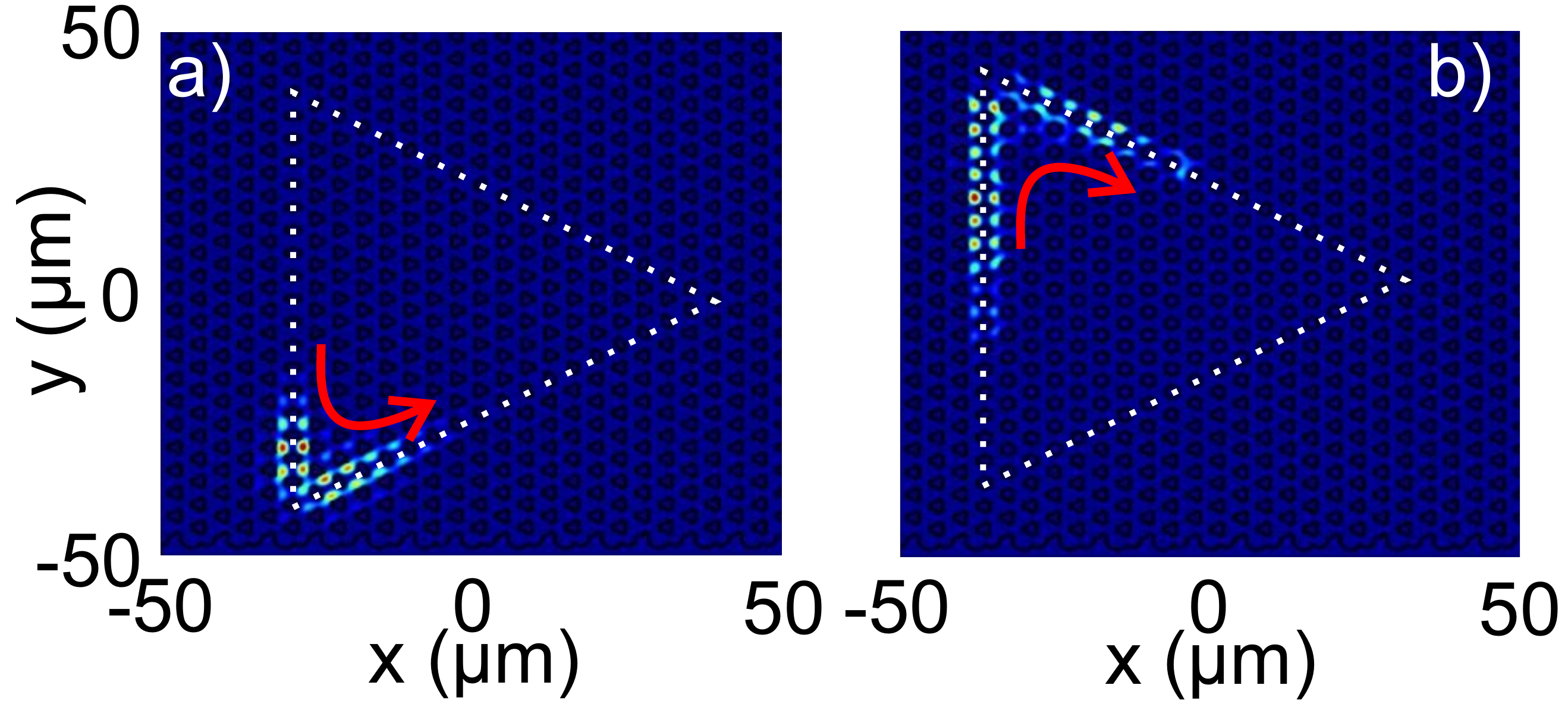}
\caption{(Color online) Behavior at the 120$^\circ$ corners of polygonal interface a) QVH/QVH b) QVH/QAH. Red arrows show the propagation direction.} 
\label{figcorners120}
 \end{figure}

Figure  \ref{figcorners120}(a) demonstrates the behavior of interface states at the 120$^\circ$ corners (meaning that the propagation direction changes by 120$^\circ$) of a polygonal interface between two mutually inverted TMD analog lattices. Such angle allows to have two interfaces of the same type (A-A and A-A or B-B and B-B). As can be seen from the figure and from the associated movie (see \cite{video1}), there is no backscattering on such interface and the wavepacket, initially created at the interface states, continues its propagation without being perturbed by such corner. Qualitatively, such corner acts like a mirror (formed by the periodic lattice), which redirects the wavepacket from the input interface state into the output interface state. Both propagation directions along the triangle are available by the choice of the initial valley excitation. Panel (b) shows the same configuration, but for an interface between the QVH/QAH phases. Here, no backscattering is possible for any type of corner, as we shall see below, because of the topological protection. The propagation direction is not defined by the excitation conditions (a single pillar of the interface is excited), but by the sign of the Zeeman splitting. However, a large gap is more difficult to obtain in the QAH phase \cite{bleu2017photonic}, and therefore one can expect to have a larger excitation of the bulk in this case. As can be seen from the movie (see \cite{video3}), it is impossible to distinguish the topologically protected QVH/QAH phase from the QVH/QVH interface which does not provide any topological protection. The absence of backscattering in one particular experiment does not mean that the system is protected completely, as we shall see below.

Figure \ref{figcorners60} shows a zoomed image of a more interesting configuration, corresponding to a 60$^\circ$ turn of the interface. Panel (a) shows the interface between the two QVH phases. For such angle, the interface changes type (from A-A to B-B or vice versa), which leads to several visible effects. First of all, the change of the interface type, similar to the change of dimerization order in the Su-Schrieffer-Heeger model \cite{Su1980}, leads to the creation of a defect state (domain wall) between the two regions. This domain wall creates a strong backscattering for the wavepacket on the interface states, contrary to the 120$^\circ$ turn, where the backscattering was completely absent. Moreover, since the interface changes type, the nature of the state changes as well: from the "anti-bonding" state of the first interface (made of two pillars with a larger radius and therefore a lower energy), the state changes into the "bonding" state of the second interface (where there are two smaller pillars). The two red arrows mark the propagation direction for the backscattered part and the main part of the wavepacket. The propagation direction for both of them is clearly visible in the movie (see \cite{video2}).

\begin{figure}[tbp]
 \includegraphics[scale=0.75]{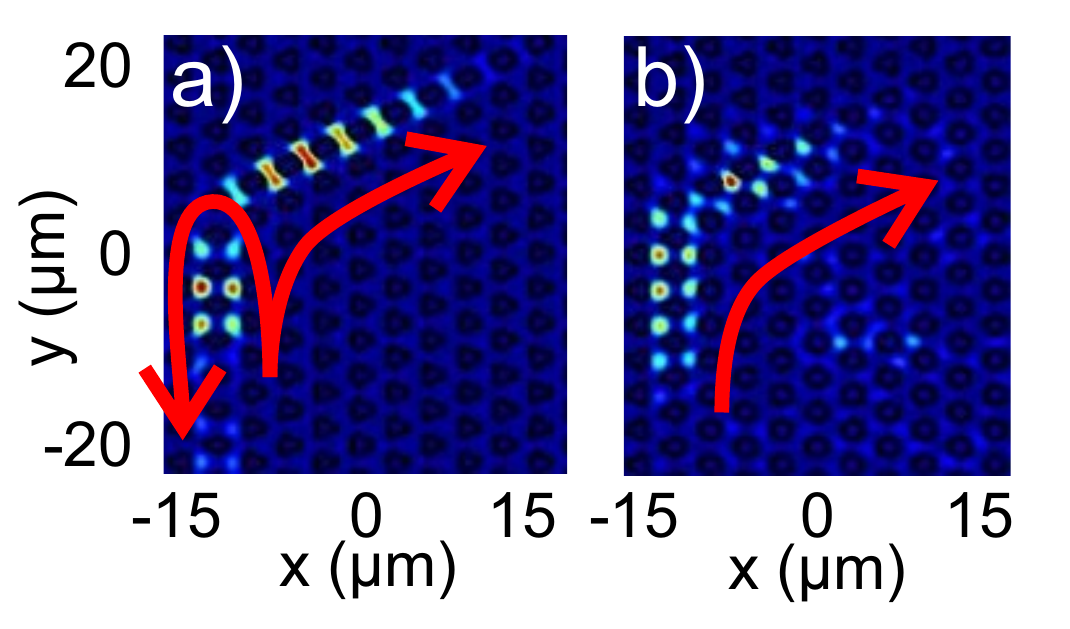}
\caption{(Color online) Conversion of the interface states at the 60$^\circ$ corners of polygonal interface a) QVH/QVH b) QVH/QAH. Red arrows show the propagation direction.} 
\label{figcorners60}
 \end{figure} 

In the QVH/QAH interface case, shown in Fig. \ref{figcorners60}(b), no backscattering is possible because of the topological protection, provided by the QAH phase, but the interface state also changes its nature at the 60$^\circ$ turn. However, because of the complicated mutual action of the TE-TM SOC and of the Zeeman splitting, it is more difficult to analyze from the spatial pattern of the total intensity. The study of such interface junctions in the perfect valley filter may be a subject for future works. While it is not so clear from the snapshot, the associated movie (see \cite{video4}) clearly shows the absence of backscattering at the junction.

\begin{figure}[tbp]
 \includegraphics[scale=0.33]{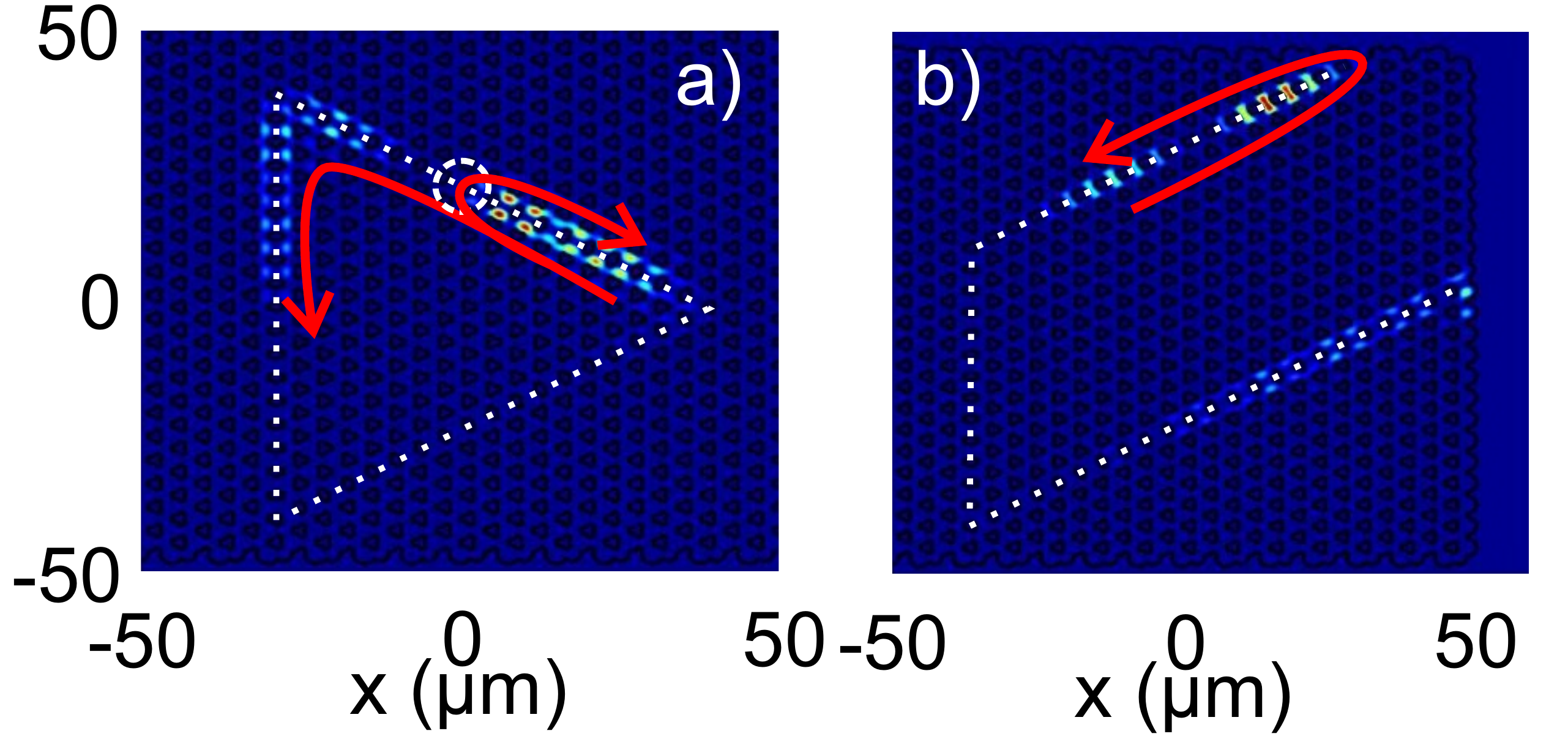}
\caption{(Color online) Sensitivity of the QVH interface states to the perturbations: backscattering a) on a localized defect b) at the boundary with the "vacuum". Red arrows show the propagation direction.} 
\label{figscatter}
 \end{figure}

Finally, we have designed a special numerical experiment to show that the QVH states indeed do not provide topological protection from backscattering. Figure \ref{figscatter} shows two distinct situations. Panel (a) demonstrates that a localized defect (created on a single pillar on the interface) leads to the backscattering of an important fraction of the wavepacket (red arrows), while at the same experiment the corners have shown no backscattering, and one might have concluded that the states are protected (see \cite{video1}). Panel (b) (which corresponds to the calculation of Fig. \ref{figcorners60} at later times, see \cite{video2}) shows that the interface states can only exist on the interface between two inverted TMD analog lattices, and they cannot propagate at the boundary between the TMD analog and the "vacuum" (absence of pillars). Such boundary leads to complete reflection of the propagating wavepacket.

 \section{Conclusion}
 In this work we highlight the analogy between electronic TMD materials and a honeycomb array of photonic pillar cavities. The control of the size of the pillars allows to control the gap in the dispersion. Valley-polarized modes are predicted at the interface between two mutually inverted TMD lattices. We study these states, taking into account the TE-TM SOC, typical for cavity systems.
We show that the control of the TE-TM SOC allows to control the number and group velocity of the interfaces modes. The observation of QVHE with light should be achievable in sufficiently regular structures.

In the second part, we propose a scheme to engineer a perfect valley filter using the interface between a QVH phase and a QAH phase predicted for polariton systems. In this configuration, the interface valley-polarized states are topologically protected against back-scattering and bulk diffusion.
Our results are supported by analytics (the computation of topological charges), tight-binding numerical dispersions, and full numerical simulations. As in several previous works \cite{Bleu2016,bleu2017photonic}, the effective Zeeman fields can be controlled by optical pumping thanks to the spin-anisotropic interactions. This opens the perspective to realize all-optically controlled, reconfigurable valley filters. Another interesting perspective is the study of the physics of interacting bosonic quantum fluids \cite{RevCarusotto2013} in topological systems \cite{Lumer2013,Engelhard2015,furukawa2015excitation,Xu2016,DiLiberto2016,Solnyshkov2016}, such as, for instance, chiral non-linear gap states \cite{Solnyshkov2017}.

We acknowledge the support of the project "Quantum Fluids of Light"  (ANR-16-CE30-0021).
 
\bibliography{biblio}

\begin{thebibliography}{60}
\expandafter\ifx\csname natexlab\endcsname\relax\def\natexlab#1{#1}\fi
\expandafter\ifx\csname bibnamefont\endcsname\relax
  \def\bibnamefont#1{#1}\fi
\expandafter\ifx\csname bibfnamefont\endcsname\relax
  \def\bibfnamefont#1{#1}\fi
\expandafter\ifx\csname citenamefont\endcsname\relax
  \def\citenamefont#1{#1}\fi
\expandafter\ifx\csname url\endcsname\relax
  \def\url#1{\texttt{#1}}\fi
\expandafter\ifx\csname urlprefix\endcsname\relax\def\urlprefix{URL }\fi
\providecommand{\bibinfo}[2]{#2}
\providecommand{\eprint}[2][]{\url{#2}}

\bibitem[{\citenamefont{Wallace}(1947)}]{Wallace1947}
\bibinfo{author}{\bibfnamefont{P.~R.} \bibnamefont{Wallace}},
  \bibinfo{journal}{Phys. Rev.} \textbf{\bibinfo{volume}{71}},
  \bibinfo{pages}{622} (\bibinfo{year}{1947}),
  \urlprefix\url{http://link.aps.org/doi/10.1103/PhysRev.71.622}.

\bibitem[{\citenamefont{Xiao et~al.}(2010)\citenamefont{Xiao, Chang, and
  Niu}}]{RevNiu2010}
\bibinfo{author}{\bibfnamefont{D.}~\bibnamefont{Xiao}},
  \bibinfo{author}{\bibfnamefont{M.-C.} \bibnamefont{Chang}}, \bibnamefont{and}
  \bibinfo{author}{\bibfnamefont{Q.}~\bibnamefont{Niu}}, \bibinfo{journal}{Rev.
  Mod. Phys.} \textbf{\bibinfo{volume}{82}}, \bibinfo{pages}{1959}
  (\bibinfo{year}{2010}),
  \urlprefix\url{http://link.aps.org/doi/10.1103/RevModPhys.82.1959}.

\bibitem[{\citenamefont{Xiao et~al.}(2007)\citenamefont{Xiao, Yao, and
  Niu}}]{Niu2007}
\bibinfo{author}{\bibfnamefont{D.}~\bibnamefont{Xiao}},
  \bibinfo{author}{\bibfnamefont{W.}~\bibnamefont{Yao}}, \bibnamefont{and}
  \bibinfo{author}{\bibfnamefont{Q.}~\bibnamefont{Niu}},
  \bibinfo{journal}{Phys. Rev. Lett.} \textbf{\bibinfo{volume}{99}},
  \bibinfo{pages}{236809} (\bibinfo{year}{2007}),
  \urlprefix\url{http://link.aps.org/doi/10.1103/PhysRevLett.99.236809}.

\bibitem[{\citenamefont{Xiao et~al.}(2012)\citenamefont{Xiao, Liu, Feng, Xu,
  and Yao}}]{Yao2012}
\bibinfo{author}{\bibfnamefont{D.}~\bibnamefont{Xiao}},
  \bibinfo{author}{\bibfnamefont{G.-B.} \bibnamefont{Liu}},
  \bibinfo{author}{\bibfnamefont{W.}~\bibnamefont{Feng}},
  \bibinfo{author}{\bibfnamefont{X.}~\bibnamefont{Xu}}, \bibnamefont{and}
  \bibinfo{author}{\bibfnamefont{W.}~\bibnamefont{Yao}},
  \bibinfo{journal}{Phys. Rev. Lett.} \textbf{\bibinfo{volume}{108}},
  \bibinfo{pages}{196802} (\bibinfo{year}{2012}),
  \urlprefix\url{http://link.aps.org/doi/10.1103/PhysRevLett.108.196802}.

\bibitem[{\citenamefont{Trushin et~al.}(2016)\citenamefont{Trushin, Goerbig,
  and Belzig}}]{Trushin2016}
\bibinfo{author}{\bibfnamefont{M.}~\bibnamefont{Trushin}},
  \bibinfo{author}{\bibfnamefont{M.~O.} \bibnamefont{Goerbig}},
  \bibnamefont{and} \bibinfo{author}{\bibfnamefont{W.}~\bibnamefont{Belzig}},
  \bibinfo{journal}{Phys. Rev. B} \textbf{\bibinfo{volume}{94}},
  \bibinfo{pages}{041301} (\bibinfo{year}{2016}),
  \urlprefix\url{http://link.aps.org/doi/10.1103/PhysRevB.94.041301}.

\bibitem[{\citenamefont{Zeng et~al.}(2012)\citenamefont{Zeng, Dai, Yao, Xiao,
  and Cui}}]{zeng2012valley}
\bibinfo{author}{\bibfnamefont{H.}~\bibnamefont{Zeng}},
  \bibinfo{author}{\bibfnamefont{J.}~\bibnamefont{Dai}},
  \bibinfo{author}{\bibfnamefont{W.}~\bibnamefont{Yao}},
  \bibinfo{author}{\bibfnamefont{D.}~\bibnamefont{Xiao}}, \bibnamefont{and}
  \bibinfo{author}{\bibfnamefont{X.}~\bibnamefont{Cui}},
  \bibinfo{journal}{Nature nanotechnology} \textbf{\bibinfo{volume}{7}},
  \bibinfo{pages}{490} (\bibinfo{year}{2012}).

\bibitem[{\citenamefont{Mak et~al.}(2012)\citenamefont{Mak, He, Shan, and
  Heinz}}]{mak2012control}
\bibinfo{author}{\bibfnamefont{K.~F.} \bibnamefont{Mak}},
  \bibinfo{author}{\bibfnamefont{K.}~\bibnamefont{He}},
  \bibinfo{author}{\bibfnamefont{J.}~\bibnamefont{Shan}}, \bibnamefont{and}
  \bibinfo{author}{\bibfnamefont{T.~F.} \bibnamefont{Heinz}},
  \bibinfo{journal}{Nature nanotechnology} \textbf{\bibinfo{volume}{7}},
  \bibinfo{pages}{494} (\bibinfo{year}{2012}).

\bibitem[{\citenamefont{Yao et~al.}(2008)\citenamefont{Yao, Xiao, and
  Niu}}]{Niu2008}
\bibinfo{author}{\bibfnamefont{W.}~\bibnamefont{Yao}},
  \bibinfo{author}{\bibfnamefont{D.}~\bibnamefont{Xiao}}, \bibnamefont{and}
  \bibinfo{author}{\bibfnamefont{Q.}~\bibnamefont{Niu}},
  \bibinfo{journal}{Phys. Rev. B} \textbf{\bibinfo{volume}{77}},
  \bibinfo{pages}{235406} (\bibinfo{year}{2008}),
  \urlprefix\url{http://link.aps.org/doi/10.1103/PhysRevB.77.235406}.

\bibitem[{\citenamefont{Cao et~al.}(2012)\citenamefont{Cao, Wang, Han, Ye, Zhu,
  Shi, Niu, Tan, Wang, Liu et~al.}}]{cao2012valley}
\bibinfo{author}{\bibfnamefont{T.}~\bibnamefont{Cao}},
  \bibinfo{author}{\bibfnamefont{G.}~\bibnamefont{Wang}},
  \bibinfo{author}{\bibfnamefont{W.}~\bibnamefont{Han}},
  \bibinfo{author}{\bibfnamefont{H.}~\bibnamefont{Ye}},
  \bibinfo{author}{\bibfnamefont{C.}~\bibnamefont{Zhu}},
  \bibinfo{author}{\bibfnamefont{J.}~\bibnamefont{Shi}},
  \bibinfo{author}{\bibfnamefont{Q.}~\bibnamefont{Niu}},
  \bibinfo{author}{\bibfnamefont{P.}~\bibnamefont{Tan}},
  \bibinfo{author}{\bibfnamefont{E.}~\bibnamefont{Wang}},
  \bibinfo{author}{\bibfnamefont{B.}~\bibnamefont{Liu}}, \bibnamefont{et~al.},
  \bibinfo{journal}{Nature communications} \textbf{\bibinfo{volume}{3}},
  \bibinfo{pages}{887} (\bibinfo{year}{2012}).

\bibitem[{\citenamefont{Mak et~al.}(2014)\citenamefont{Mak, McGill, Park, and
  McEuen}}]{mak2014valley}
\bibinfo{author}{\bibfnamefont{K.~F.} \bibnamefont{Mak}},
  \bibinfo{author}{\bibfnamefont{K.~L.} \bibnamefont{McGill}},
  \bibinfo{author}{\bibfnamefont{J.}~\bibnamefont{Park}}, \bibnamefont{and}
  \bibinfo{author}{\bibfnamefont{P.~L.} \bibnamefont{McEuen}},
  \bibinfo{journal}{Science} \textbf{\bibinfo{volume}{344}},
  \bibinfo{pages}{1489} (\bibinfo{year}{2014}).

\bibitem[{\citenamefont{Kane and Mele}(2005)}]{kane2005z}
\bibinfo{author}{\bibfnamefont{C.~L.} \bibnamefont{Kane}} \bibnamefont{and}
  \bibinfo{author}{\bibfnamefont{E.~J.} \bibnamefont{Mele}},
  \bibinfo{journal}{Physical review letters} \textbf{\bibinfo{volume}{95}},
  \bibinfo{pages}{146802} (\bibinfo{year}{2005}).

\bibitem[{\citenamefont{Semenoff et~al.}(2008)\citenamefont{Semenoff, Semenoff,
  and Zhou}}]{Semenoff2008}
\bibinfo{author}{\bibfnamefont{G.~W.} \bibnamefont{Semenoff}},
  \bibinfo{author}{\bibfnamefont{V.}~\bibnamefont{Semenoff}}, \bibnamefont{and}
  \bibinfo{author}{\bibfnamefont{F.}~\bibnamefont{Zhou}},
  \bibinfo{journal}{Phys. Rev. Lett.} \textbf{\bibinfo{volume}{101}},
  \bibinfo{pages}{087204} (\bibinfo{year}{2008}),
  \urlprefix\url{http://link.aps.org/doi/10.1103/PhysRevLett.101.087204}.

\bibitem[{\citenamefont{Yao et~al.}(2009)\citenamefont{Yao, Yang, and
  Niu}}]{Niu2009}
\bibinfo{author}{\bibfnamefont{W.}~\bibnamefont{Yao}},
  \bibinfo{author}{\bibfnamefont{S.~A.} \bibnamefont{Yang}}, \bibnamefont{and}
  \bibinfo{author}{\bibfnamefont{Q.}~\bibnamefont{Niu}},
  \bibinfo{journal}{Phys. Rev. Lett.} \textbf{\bibinfo{volume}{102}},
  \bibinfo{pages}{096801} (\bibinfo{year}{2009}),
  \urlprefix\url{http://link.aps.org/doi/10.1103/PhysRevLett.102.096801}.

\bibitem[{\citenamefont{McCann}(2006)}]{McCann2006}
\bibinfo{author}{\bibfnamefont{E.}~\bibnamefont{McCann}},
  \bibinfo{journal}{Phys. Rev. B} \textbf{\bibinfo{volume}{74}},
  \bibinfo{pages}{161403} (\bibinfo{year}{2006}),
  \urlprefix\url{http://link.aps.org/doi/10.1103/PhysRevB.74.161403}.

\bibitem[{\citenamefont{Castro et~al.}(2007)\citenamefont{Castro, Novoselov,
  Morozov, Peres, dos Santos, Nilsson, Guinea, Geim, and
  Neto}}]{GeimCastro2007}
\bibinfo{author}{\bibfnamefont{E.~V.} \bibnamefont{Castro}},
  \bibinfo{author}{\bibfnamefont{K.~S.} \bibnamefont{Novoselov}},
  \bibinfo{author}{\bibfnamefont{S.~V.} \bibnamefont{Morozov}},
  \bibinfo{author}{\bibfnamefont{N.~M.~R.} \bibnamefont{Peres}},
  \bibinfo{author}{\bibfnamefont{J.~M. B.~L.} \bibnamefont{dos Santos}},
  \bibinfo{author}{\bibfnamefont{J.}~\bibnamefont{Nilsson}},
  \bibinfo{author}{\bibfnamefont{F.}~\bibnamefont{Guinea}},
  \bibinfo{author}{\bibfnamefont{A.~K.} \bibnamefont{Geim}}, \bibnamefont{and}
  \bibinfo{author}{\bibfnamefont{A.~H.~C.} \bibnamefont{Neto}},
  \bibinfo{journal}{Phys. Rev. Lett.} \textbf{\bibinfo{volume}{99}},
  \bibinfo{pages}{216802} (\bibinfo{year}{2007}),
  \urlprefix\url{http://link.aps.org/doi/10.1103/PhysRevLett.99.216802}.

\bibitem[{\citenamefont{Martin et~al.}(2008)\citenamefont{Martin, Blanter, and
  Morpurgo}}]{Martin2008}
\bibinfo{author}{\bibfnamefont{I.}~\bibnamefont{Martin}},
  \bibinfo{author}{\bibfnamefont{Y.~M.} \bibnamefont{Blanter}},
  \bibnamefont{and} \bibinfo{author}{\bibfnamefont{A.~F.}
  \bibnamefont{Morpurgo}}, \bibinfo{journal}{Phys. Rev. Lett.}
  \textbf{\bibinfo{volume}{100}}, \bibinfo{pages}{036804}
  (\bibinfo{year}{2008}),
  \urlprefix\url{http://link.aps.org/doi/10.1103/PhysRevLett.100.036804}.

\bibitem[{\citenamefont{Li et~al.}(2011)\citenamefont{Li, Martin, B{\"u}ttiker,
  and Morpurgo}}]{li2011topological}
\bibinfo{author}{\bibfnamefont{J.}~\bibnamefont{Li}},
  \bibinfo{author}{\bibfnamefont{I.}~\bibnamefont{Martin}},
  \bibinfo{author}{\bibfnamefont{M.}~\bibnamefont{B{\"u}ttiker}},
  \bibnamefont{and} \bibinfo{author}{\bibfnamefont{A.~F.}
  \bibnamefont{Morpurgo}}, \bibinfo{journal}{Nature Physics}
  \textbf{\bibinfo{volume}{7}}, \bibinfo{pages}{38} (\bibinfo{year}{2011}).

\bibitem[{\citenamefont{Zhang et~al.}(2013{\natexlab{a}})\citenamefont{Zhang,
  MacDonald, and Mele}}]{zhang2013valley}
\bibinfo{author}{\bibfnamefont{F.}~\bibnamefont{Zhang}},
  \bibinfo{author}{\bibfnamefont{A.~H.} \bibnamefont{MacDonald}},
  \bibnamefont{and} \bibinfo{author}{\bibfnamefont{E.~J.} \bibnamefont{Mele}},
  \bibinfo{journal}{Proceedings of the National Academy of Sciences}
  \textbf{\bibinfo{volume}{110}}, \bibinfo{pages}{10546}
  (\bibinfo{year}{2013}{\natexlab{a}}).

\bibitem[{\citenamefont{Castro et~al.}(2008)\citenamefont{Castro, Peres,
  Lopes~dos Santos, Neto, and Guinea}}]{Castro2008}
\bibinfo{author}{\bibfnamefont{E.~V.} \bibnamefont{Castro}},
  \bibinfo{author}{\bibfnamefont{N.~M.~R.} \bibnamefont{Peres}},
  \bibinfo{author}{\bibfnamefont{J.~M.~B.} \bibnamefont{Lopes~dos Santos}},
  \bibinfo{author}{\bibfnamefont{A.~H.~C.} \bibnamefont{Neto}},
  \bibnamefont{and} \bibinfo{author}{\bibfnamefont{F.}~\bibnamefont{Guinea}},
  \bibinfo{journal}{Phys. Rev. Lett.} \textbf{\bibinfo{volume}{100}},
  \bibinfo{pages}{026802} (\bibinfo{year}{2008}),
  \urlprefix\url{http://link.aps.org/doi/10.1103/PhysRevLett.100.026802}.

\bibitem[{\citenamefont{Ju et~al.}(2015)\citenamefont{Ju, Shi, Nair, Lv, Jin,
  Velasco~Jr, Ojeda-Aristizabal, Bechtel, Martin, Zettl
  et~al.}}]{ju2015topological}
\bibinfo{author}{\bibfnamefont{L.}~\bibnamefont{Ju}},
  \bibinfo{author}{\bibfnamefont{Z.}~\bibnamefont{Shi}},
  \bibinfo{author}{\bibfnamefont{N.}~\bibnamefont{Nair}},
  \bibinfo{author}{\bibfnamefont{Y.}~\bibnamefont{Lv}},
  \bibinfo{author}{\bibfnamefont{C.}~\bibnamefont{Jin}},
  \bibinfo{author}{\bibfnamefont{J.}~\bibnamefont{Velasco~Jr}},
  \bibinfo{author}{\bibfnamefont{C.}~\bibnamefont{Ojeda-Aristizabal}},
  \bibinfo{author}{\bibfnamefont{H.~A.} \bibnamefont{Bechtel}},
  \bibinfo{author}{\bibfnamefont{M.~C.} \bibnamefont{Martin}},
  \bibinfo{author}{\bibfnamefont{A.}~\bibnamefont{Zettl}},
  \bibnamefont{et~al.}, \bibinfo{journal}{Nature}
  \textbf{\bibinfo{volume}{520}}, \bibinfo{pages}{650} (\bibinfo{year}{2015}).

\bibitem[{\citenamefont{Vaezi et~al.}(2013)\citenamefont{Vaezi, Liang, Ngai,
  Yang, and Kim}}]{Vaezi2013}
\bibinfo{author}{\bibfnamefont{A.}~\bibnamefont{Vaezi}},
  \bibinfo{author}{\bibfnamefont{Y.}~\bibnamefont{Liang}},
  \bibinfo{author}{\bibfnamefont{D.~H.} \bibnamefont{Ngai}},
  \bibinfo{author}{\bibfnamefont{L.}~\bibnamefont{Yang}}, \bibnamefont{and}
  \bibinfo{author}{\bibfnamefont{E.-A.} \bibnamefont{Kim}},
  \bibinfo{journal}{Phys. Rev. X} \textbf{\bibinfo{volume}{3}},
  \bibinfo{pages}{021018} (\bibinfo{year}{2013}),
  \urlprefix\url{http://link.aps.org/doi/10.1103/PhysRevX.3.021018}.

\bibitem[{\citenamefont{Marino et~al.}(2015)\citenamefont{Marino, Nascimento,
  Alves, and Smith}}]{Marino2015}
\bibinfo{author}{\bibfnamefont{E.~C.} \bibnamefont{Marino}},
  \bibinfo{author}{\bibfnamefont{L.~O.} \bibnamefont{Nascimento}},
  \bibinfo{author}{\bibfnamefont{V.~S.} \bibnamefont{Alves}}, \bibnamefont{and}
  \bibinfo{author}{\bibfnamefont{C.~M.} \bibnamefont{Smith}},
  \bibinfo{journal}{Phys. Rev. X} \textbf{\bibinfo{volume}{5}},
  \bibinfo{pages}{011040} (\bibinfo{year}{2015}),
  \urlprefix\url{http://link.aps.org/doi/10.1103/PhysRevX.5.011040}.

\bibitem[{\citenamefont{Ren et~al.}(2016)\citenamefont{Ren, Qiao, and
  Niu}}]{ren2016topological}
\bibinfo{author}{\bibfnamefont{Y.}~\bibnamefont{Ren}},
  \bibinfo{author}{\bibfnamefont{Z.}~\bibnamefont{Qiao}}, \bibnamefont{and}
  \bibinfo{author}{\bibfnamefont{Q.}~\bibnamefont{Niu}},
  \bibinfo{journal}{Reports on Progress in Physics}
  \textbf{\bibinfo{volume}{79}}, \bibinfo{pages}{066501}
  (\bibinfo{year}{2016}).

\bibitem[{\citenamefont{Pan et~al.}(2014)\citenamefont{Pan, Li, Liu, Zhu, Qiao,
  and Yao}}]{Yao2014}
\bibinfo{author}{\bibfnamefont{H.}~\bibnamefont{Pan}},
  \bibinfo{author}{\bibfnamefont{Z.}~\bibnamefont{Li}},
  \bibinfo{author}{\bibfnamefont{C.-C.} \bibnamefont{Liu}},
  \bibinfo{author}{\bibfnamefont{G.}~\bibnamefont{Zhu}},
  \bibinfo{author}{\bibfnamefont{Z.}~\bibnamefont{Qiao}}, \bibnamefont{and}
  \bibinfo{author}{\bibfnamefont{Y.}~\bibnamefont{Yao}},
  \bibinfo{journal}{Phys. Rev. Lett.} \textbf{\bibinfo{volume}{112}},
  \bibinfo{pages}{106802} (\bibinfo{year}{2014}),
  \urlprefix\url{http://link.aps.org/doi/10.1103/PhysRevLett.112.106802}.

\bibitem[{\citenamefont{Ozawa and Carusotto}(2014)}]{Carusotto2014}
\bibinfo{author}{\bibfnamefont{T.}~\bibnamefont{Ozawa}} \bibnamefont{and}
  \bibinfo{author}{\bibfnamefont{I.}~\bibnamefont{Carusotto}},
  \bibinfo{journal}{Phys. Rev. Lett.} \textbf{\bibinfo{volume}{112}},
  \bibinfo{pages}{133902} (\bibinfo{year}{2014}),
  \urlprefix\url{http://link.aps.org/doi/10.1103/PhysRevLett.112.133902}.

\bibitem[{\citenamefont{Barik et~al.}(2016)\citenamefont{Barik, Miyake,
  DeGottardi, Waks, and Hafezi}}]{barik2016two}
\bibinfo{author}{\bibfnamefont{S.}~\bibnamefont{Barik}},
  \bibinfo{author}{\bibfnamefont{H.}~\bibnamefont{Miyake}},
  \bibinfo{author}{\bibfnamefont{W.}~\bibnamefont{DeGottardi}},
  \bibinfo{author}{\bibfnamefont{E.}~\bibnamefont{Waks}}, \bibnamefont{and}
  \bibinfo{author}{\bibfnamefont{M.}~\bibnamefont{Hafezi}},
  \bibinfo{journal}{New Journal of Physics} \textbf{\bibinfo{volume}{18}},
  \bibinfo{pages}{113013} (\bibinfo{year}{2016}).

\bibitem[{\citenamefont{Ma et~al.}(2015)\citenamefont{Ma, Khanikaev, Mousavi,
  and Shvets}}]{Shvets2015}
\bibinfo{author}{\bibfnamefont{T.}~\bibnamefont{Ma}},
  \bibinfo{author}{\bibfnamefont{A.~B.} \bibnamefont{Khanikaev}},
  \bibinfo{author}{\bibfnamefont{S.~H.} \bibnamefont{Mousavi}},
  \bibnamefont{and} \bibinfo{author}{\bibfnamefont{G.}~\bibnamefont{Shvets}},
  \bibinfo{journal}{Phys. Rev. Lett.} \textbf{\bibinfo{volume}{114}},
  \bibinfo{pages}{127401} (\bibinfo{year}{2015}),
  \urlprefix\url{http://link.aps.org/doi/10.1103/PhysRevLett.114.127401}.

\bibitem[{\citenamefont{Ma and Shvets}(2016)}]{ma2016all}
\bibinfo{author}{\bibfnamefont{T.}~\bibnamefont{Ma}} \bibnamefont{and}
  \bibinfo{author}{\bibfnamefont{G.}~\bibnamefont{Shvets}},
  \bibinfo{journal}{New Journal of Physics} \textbf{\bibinfo{volume}{18}},
  \bibinfo{pages}{025012} (\bibinfo{year}{2016}).

\bibitem[{\citenamefont{Chen and Dong}(2016)}]{chen2016valley}
\bibinfo{author}{\bibfnamefont{X.-D.} \bibnamefont{Chen}} \bibnamefont{and}
  \bibinfo{author}{\bibfnamefont{J.-W.} \bibnamefont{Dong}},
  \bibinfo{journal}{arXiv preprint arXiv:1602.03352}  (\bibinfo{year}{2016}).

\bibitem[{\citenamefont{Jacqmin et~al.}(2014)\citenamefont{Jacqmin, Carusotto,
  Sagnes, Abbarchi, Solnyshkov, Malpuech, Galopin, Lema\^{\i}tre, Bloch, and
  Amo}}]{Jacqmin2014}
\bibinfo{author}{\bibfnamefont{T.}~\bibnamefont{Jacqmin}},
  \bibinfo{author}{\bibfnamefont{I.}~\bibnamefont{Carusotto}},
  \bibinfo{author}{\bibfnamefont{I.}~\bibnamefont{Sagnes}},
  \bibinfo{author}{\bibfnamefont{M.}~\bibnamefont{Abbarchi}},
  \bibinfo{author}{\bibfnamefont{D.~D.} \bibnamefont{Solnyshkov}},
  \bibinfo{author}{\bibfnamefont{G.}~\bibnamefont{Malpuech}},
  \bibinfo{author}{\bibfnamefont{E.}~\bibnamefont{Galopin}},
  \bibinfo{author}{\bibfnamefont{A.}~\bibnamefont{Lema\^{\i}tre}},
  \bibinfo{author}{\bibfnamefont{J.}~\bibnamefont{Bloch}}, \bibnamefont{and}
  \bibinfo{author}{\bibfnamefont{A.}~\bibnamefont{Amo}},
  \bibinfo{journal}{Phys. Rev. Lett.} \textbf{\bibinfo{volume}{112}},
  \bibinfo{pages}{116402} (\bibinfo{year}{2014}),
  \urlprefix\url{http://link.aps.org/doi/10.1103/PhysRevLett.112.116402}.

\bibitem[{\citenamefont{Nalitov
  et~al.}(2015{\natexlab{a}})\citenamefont{Nalitov, Solnyshkov, and
  Malpuech}}]{polaritonZ2015}
\bibinfo{author}{\bibfnamefont{A.~V.} \bibnamefont{Nalitov}},
  \bibinfo{author}{\bibfnamefont{D.~D.} \bibnamefont{Solnyshkov}},
  \bibnamefont{and} \bibinfo{author}{\bibfnamefont{G.}~\bibnamefont{Malpuech}},
  \bibinfo{journal}{Phys. Rev. Lett.} \textbf{\bibinfo{volume}{114}},
  \bibinfo{pages}{116401} (\bibinfo{year}{2015}{\natexlab{a}}),
  \urlprefix\url{http://link.aps.org/doi/10.1103/PhysRevLett.114.116401}.

\bibitem[{\citenamefont{Karzig et~al.}(2015)\citenamefont{Karzig, Bardyn,
  Lindner, and Refael}}]{KarzigPRX2015}
\bibinfo{author}{\bibfnamefont{T.}~\bibnamefont{Karzig}},
  \bibinfo{author}{\bibfnamefont{C.-E.} \bibnamefont{Bardyn}},
  \bibinfo{author}{\bibfnamefont{N.~H.} \bibnamefont{Lindner}},
  \bibnamefont{and} \bibinfo{author}{\bibfnamefont{G.}~\bibnamefont{Refael}},
  \bibinfo{journal}{Phys. Rev. X} \textbf{\bibinfo{volume}{5}},
  \bibinfo{pages}{031001} (\bibinfo{year}{2015}),
  \urlprefix\url{http://link.aps.org/doi/10.1103/PhysRevX.5.031001}.

\bibitem[{\citenamefont{Bardyn et~al.}(2015)\citenamefont{Bardyn, Karzig,
  Refael, and Liew}}]{LiewPRB2015}
\bibinfo{author}{\bibfnamefont{C.-E.} \bibnamefont{Bardyn}},
  \bibinfo{author}{\bibfnamefont{T.}~\bibnamefont{Karzig}},
  \bibinfo{author}{\bibfnamefont{G.}~\bibnamefont{Refael}}, \bibnamefont{and}
  \bibinfo{author}{\bibfnamefont{T.~C.~H.} \bibnamefont{Liew}},
  \bibinfo{journal}{Phys. Rev. B} \textbf{\bibinfo{volume}{91}},
  \bibinfo{pages}{161413} (\bibinfo{year}{2015}),
  \urlprefix\url{http://link.aps.org/doi/10.1103/PhysRevB.91.161413}.

\bibitem[{\citenamefont{Gulevich et~al.}(2016)\citenamefont{Gulevich, Yudin,
  Iorsh, and Shelykh}}]{gulevich2016kagome}
\bibinfo{author}{\bibfnamefont{D.~R.} \bibnamefont{Gulevich}},
  \bibinfo{author}{\bibfnamefont{D.}~\bibnamefont{Yudin}},
  \bibinfo{author}{\bibfnamefont{I.~V.} \bibnamefont{Iorsh}}, \bibnamefont{and}
  \bibinfo{author}{\bibfnamefont{I.~A.} \bibnamefont{Shelykh}},
  \bibinfo{journal}{Phys. Rev. B} \textbf{\bibinfo{volume}{94}},
  \bibinfo{pages}{115437} (\bibinfo{year}{2016}),
  \urlprefix\url{http://link.aps.org/doi/10.1103/PhysRevB.94.115437}.

\bibitem[{\citenamefont{Kavokin et~al.}(2011)\citenamefont{Kavokin, Baumberg,
  Malpuech, and Laussy}}]{kavokin2011microcavities}
\bibinfo{author}{\bibfnamefont{A.}~\bibnamefont{Kavokin}},
  \bibinfo{author}{\bibfnamefont{J.~J.} \bibnamefont{Baumberg}},
  \bibinfo{author}{\bibfnamefont{G.}~\bibnamefont{Malpuech}}, \bibnamefont{and}
  \bibinfo{author}{\bibfnamefont{F.~P.} \bibnamefont{Laussy}},
  \emph{\bibinfo{title}{Microcavities}}, vol.~\bibinfo{volume}{16}
  (\bibinfo{publisher}{OUP Oxford}, \bibinfo{year}{2011}).

\bibitem[{\citenamefont{Shelykh et~al.}(2009)\citenamefont{Shelykh, Kavokin,
  Rubo, Liew, and Malpuech}}]{shelykh2009polariton}
\bibinfo{author}{\bibfnamefont{I.}~\bibnamefont{Shelykh}},
  \bibinfo{author}{\bibfnamefont{A.}~\bibnamefont{Kavokin}},
  \bibinfo{author}{\bibfnamefont{Y.~G.} \bibnamefont{Rubo}},
  \bibinfo{author}{\bibfnamefont{T.}~\bibnamefont{Liew}}, \bibnamefont{and}
  \bibinfo{author}{\bibfnamefont{G.}~\bibnamefont{Malpuech}},
  \bibinfo{journal}{Semiconductor Science and Technology}
  \textbf{\bibinfo{volume}{25}}, \bibinfo{pages}{013001}
  (\bibinfo{year}{2009}).

\bibitem[{\citenamefont{Mili{\'c}evi{\'c}
  et~al.}(2015)\citenamefont{Mili{\'c}evi{\'c}, Ozawa, Andreakou, Carusotto,
  Jacqmin, Galopin, Lema{\^\i}tre, Le~Gratiet, Sagnes, Bloch
  et~al.}}]{milicevic2015edge}
\bibinfo{author}{\bibfnamefont{M.}~\bibnamefont{Mili{\'c}evi{\'c}}},
  \bibinfo{author}{\bibfnamefont{T.}~\bibnamefont{Ozawa}},
  \bibinfo{author}{\bibfnamefont{P.}~\bibnamefont{Andreakou}},
  \bibinfo{author}{\bibfnamefont{I.}~\bibnamefont{Carusotto}},
  \bibinfo{author}{\bibfnamefont{T.}~\bibnamefont{Jacqmin}},
  \bibinfo{author}{\bibfnamefont{E.}~\bibnamefont{Galopin}},
  \bibinfo{author}{\bibfnamefont{A.}~\bibnamefont{Lema{\^\i}tre}},
  \bibinfo{author}{\bibfnamefont{L.}~\bibnamefont{Le~Gratiet}},
  \bibinfo{author}{\bibfnamefont{I.}~\bibnamefont{Sagnes}},
  \bibinfo{author}{\bibfnamefont{J.}~\bibnamefont{Bloch}},
  \bibnamefont{et~al.}, \bibinfo{journal}{2D Materials}
  \textbf{\bibinfo{volume}{2}}, \bibinfo{pages}{034012} (\bibinfo{year}{2015}).

\bibitem[{\citenamefont{Zhang et~al.}(2013{\natexlab{b}})\citenamefont{Zhang,
  Qiao, and Sun}}]{Feng2013}
\bibinfo{author}{\bibfnamefont{Y.-T.} \bibnamefont{Zhang}},
  \bibinfo{author}{\bibfnamefont{Z.}~\bibnamefont{Qiao}}, \bibnamefont{and}
  \bibinfo{author}{\bibfnamefont{Q.-F.} \bibnamefont{Sun}},
  \bibinfo{journal}{Phys. Rev. B} \textbf{\bibinfo{volume}{87}},
  \bibinfo{pages}{235405} (\bibinfo{year}{2013}{\natexlab{b}}),
  \urlprefix\url{http://link.aps.org/doi/10.1103/PhysRevB.87.235405}.

\bibitem[{\citenamefont{Li et~al.}(2010)\citenamefont{Li, Morpurgo, B\"uttiker,
  and Martin}}]{Li2010}
\bibinfo{author}{\bibfnamefont{J.}~\bibnamefont{Li}},
  \bibinfo{author}{\bibfnamefont{A.~F.} \bibnamefont{Morpurgo}},
  \bibinfo{author}{\bibfnamefont{M.}~\bibnamefont{B\"uttiker}},
  \bibnamefont{and} \bibinfo{author}{\bibfnamefont{I.}~\bibnamefont{Martin}},
  \bibinfo{journal}{Phys. Rev. B} \textbf{\bibinfo{volume}{82}},
  \bibinfo{pages}{245404} (\bibinfo{year}{2010}),
  \urlprefix\url{http://link.aps.org/doi/10.1103/PhysRevB.82.245404}.

\bibitem[{\citenamefont{Volovik}(2003)}]{volovik2003universe}
\bibinfo{author}{\bibfnamefont{G.~E.} \bibnamefont{Volovik}},
  \emph{\bibinfo{title}{The universe in a helium droplet}}, vol.
  \bibinfo{volume}{117} (\bibinfo{publisher}{Oxford University Press on
  Demand}, \bibinfo{year}{2003}).

\bibitem[{\citenamefont{Solnyshkov and Malpuech}(2016)}]{CRAS2016}
\bibinfo{author}{\bibfnamefont{D.}~\bibnamefont{Solnyshkov}} \bibnamefont{and}
  \bibinfo{author}{\bibfnamefont{G.}~\bibnamefont{Malpuech}},
  \bibinfo{journal}{Comptes Rendus Physique} \textbf{\bibinfo{volume}{17}},
  \bibinfo{pages}{920} (\bibinfo{year}{2016}).

\bibitem[{\citenamefont{Nalitov
  et~al.}(2015{\natexlab{b}})\citenamefont{Nalitov, Malpuech,
  Ter\ifmmode~\mbox{\c{c}}\else \c{c}\fi{}as, and Solnyshkov}}]{Anton2015}
\bibinfo{author}{\bibfnamefont{A.~V.} \bibnamefont{Nalitov}},
  \bibinfo{author}{\bibfnamefont{G.}~\bibnamefont{Malpuech}},
  \bibinfo{author}{\bibfnamefont{H.}~\bibnamefont{Ter\ifmmode~\mbox{\c{c}}\else
  \c{c}\fi{}as}}, \bibnamefont{and} \bibinfo{author}{\bibfnamefont{D.~D.}
  \bibnamefont{Solnyshkov}}, \bibinfo{journal}{Phys. Rev. Lett.}
  \textbf{\bibinfo{volume}{114}}, \bibinfo{pages}{026803}
  (\bibinfo{year}{2015}{\natexlab{b}}),
  \urlprefix\url{http://link.aps.org/doi/10.1103/PhysRevLett.114.026803}.

\bibitem[{\citenamefont{Bleu et~al.}(2016{\natexlab{a}})\citenamefont{Bleu,
  Malpuech, and Solnyshkov}}]{bleu2016effective}
\bibinfo{author}{\bibfnamefont{O.}~\bibnamefont{Bleu}},
  \bibinfo{author}{\bibfnamefont{G.}~\bibnamefont{Malpuech}}, \bibnamefont{and}
  \bibinfo{author}{\bibfnamefont{D.}~\bibnamefont{Solnyshkov}},
  \bibinfo{journal}{arXiv preprint arXiv:1612.02998}
  (\bibinfo{year}{2016}{\natexlab{a}}).

\bibitem[{\citenamefont{Bleu et~al.}(2017)\citenamefont{Bleu, Solnyshkov, and
  Malpuech}}]{bleu2017photonic}
\bibinfo{author}{\bibfnamefont{O.}~\bibnamefont{Bleu}},
  \bibinfo{author}{\bibfnamefont{D.~D.} \bibnamefont{Solnyshkov}},
  \bibnamefont{and} \bibinfo{author}{\bibfnamefont{G.}~\bibnamefont{Malpuech}},
  \bibinfo{journal}{Phys. Rev. B} \textbf{\bibinfo{volume}{95}},
  \bibinfo{pages}{115415} (\bibinfo{year}{2017}),
  \urlprefix\url{http://link.aps.org/doi/10.1103/PhysRevB.95.115415}.

\bibitem[{\citenamefont{Pan et~al.}(2015)\citenamefont{Pan, Li, Zhang, and
  Yang}}]{Pan2015}
\bibinfo{author}{\bibfnamefont{H.}~\bibnamefont{Pan}},
  \bibinfo{author}{\bibfnamefont{X.}~\bibnamefont{Li}},
  \bibinfo{author}{\bibfnamefont{F.}~\bibnamefont{Zhang}}, \bibnamefont{and}
  \bibinfo{author}{\bibfnamefont{S.~A.} \bibnamefont{Yang}},
  \bibinfo{journal}{Phys. Rev. B} \textbf{\bibinfo{volume}{92}},
  \bibinfo{pages}{041404} (\bibinfo{year}{2015}),
  \urlprefix\url{http://link.aps.org/doi/10.1103/PhysRevB.92.041404}.

\bibitem[{sup()}]{suppl}
\bibinfo{note}{See Supplemental Material at [URL will be inserted by publisher]
  for more details on the calculations.}

\bibitem[{vid({\natexlab{a}})}]{video1}
\emph{\bibinfo{title}{Video 1: Qvh interface states for an equilateral
  triangle}}, \bibinfo{howpublished}{\url{https://youtu.be/QNwNX1k6s9M}}.

\bibitem[{vid({\natexlab{b}})}]{video3}
\emph{\bibinfo{title}{Video 3: Qah states with complete topological protection
  (120 degrees turn + defect)}},
  \bibinfo{howpublished}{\url{https://youtu.be/MXsaeHvZsLQ}}.

\bibitem[{\citenamefont{Su et~al.}(1980)\citenamefont{Su, Schrieffer, and
  Heeger}}]{Su1980}
\bibinfo{author}{\bibfnamefont{W.~P.} \bibnamefont{Su}},
  \bibinfo{author}{\bibfnamefont{J.~R.} \bibnamefont{Schrieffer}},
  \bibnamefont{and} \bibinfo{author}{\bibfnamefont{A.~J.}
  \bibnamefont{Heeger}}, \bibinfo{journal}{Phys. Rev. B}
  \textbf{\bibinfo{volume}{22}}, \bibinfo{pages}{2099} (\bibinfo{year}{1980}),
  \urlprefix\url{http://link.aps.org/doi/10.1103/PhysRevB.22.2099}.

\bibitem[{vid({\natexlab{c}})}]{video2}
\emph{\bibinfo{title}{Video 2: Qvh interface states for a 60 degrees turn +
  vacuum interface}},
  \bibinfo{howpublished}{\url{https://youtu.be/Hvr0pKh-rtM}}.

\bibitem[{vid({\natexlab{d}})}]{video4}
\emph{\bibinfo{title}{Video 4: Qah states with no backscattering for a 60
  degrees turn, a defect, and an interface with the vacuum}},
  \bibinfo{howpublished}{\url{https://youtu.be/pwNJ5h8TGCU}}.

\bibitem[{\citenamefont{Bleu et~al.}(2016{\natexlab{b}})\citenamefont{Bleu,
  Solnyshkov, and Malpuech}}]{Bleu2016}
\bibinfo{author}{\bibfnamefont{O.}~\bibnamefont{Bleu}},
  \bibinfo{author}{\bibfnamefont{D.~D.} \bibnamefont{Solnyshkov}},
  \bibnamefont{and} \bibinfo{author}{\bibfnamefont{G.}~\bibnamefont{Malpuech}},
  \bibinfo{journal}{Phys. Rev. B} \textbf{\bibinfo{volume}{93}},
  \bibinfo{pages}{085438} (\bibinfo{year}{2016}{\natexlab{b}}),
  \urlprefix\url{http://link.aps.org/doi/10.1103/PhysRevB.93.085438}.

\bibitem[{\citenamefont{Carusotto and Ciuti}(2013)}]{RevCarusotto2013}
\bibinfo{author}{\bibfnamefont{I.}~\bibnamefont{Carusotto}} \bibnamefont{and}
  \bibinfo{author}{\bibfnamefont{C.}~\bibnamefont{Ciuti}},
  \bibinfo{journal}{Rev. Mod. Phys.} \textbf{\bibinfo{volume}{85}},
  \bibinfo{pages}{299} (\bibinfo{year}{2013}),
  \urlprefix\url{http://link.aps.org/doi/10.1103/RevModPhys.85.299}.

\bibitem[{\citenamefont{Lumer et~al.}(2013)\citenamefont{Lumer, Plotnik,
  Rechtsman, and Segev}}]{Lumer2013}
\bibinfo{author}{\bibfnamefont{Y.}~\bibnamefont{Lumer}},
  \bibinfo{author}{\bibfnamefont{Y.}~\bibnamefont{Plotnik}},
  \bibinfo{author}{\bibfnamefont{M.~C.} \bibnamefont{Rechtsman}},
  \bibnamefont{and} \bibinfo{author}{\bibfnamefont{M.}~\bibnamefont{Segev}},
  \bibinfo{journal}{Phys. Rev. Lett.} \textbf{\bibinfo{volume}{111}},
  \bibinfo{pages}{243905} (\bibinfo{year}{2013}),
  \urlprefix\url{http://link.aps.org/doi/10.1103/PhysRevLett.111.243905}.

\bibitem[{\citenamefont{Engelhardt and Brandes}(2015)}]{Engelhard2015}
\bibinfo{author}{\bibfnamefont{G.}~\bibnamefont{Engelhardt}} \bibnamefont{and}
  \bibinfo{author}{\bibfnamefont{T.}~\bibnamefont{Brandes}},
  \bibinfo{journal}{Phys. Rev. A} \textbf{\bibinfo{volume}{91}},
  \bibinfo{pages}{053621} (\bibinfo{year}{2015}),
  \urlprefix\url{http://link.aps.org/doi/10.1103/PhysRevA.91.053621}.

\bibitem[{\citenamefont{Furukawa and Ueda}(2015)}]{furukawa2015excitation}
\bibinfo{author}{\bibfnamefont{S.}~\bibnamefont{Furukawa}} \bibnamefont{and}
  \bibinfo{author}{\bibfnamefont{M.}~\bibnamefont{Ueda}}, \bibinfo{journal}{New
  Journal of Physics} \textbf{\bibinfo{volume}{17}}, \bibinfo{pages}{115014}
  (\bibinfo{year}{2015}).

\bibitem[{\citenamefont{Xu et~al.}(2016)\citenamefont{Xu, You, Hemmerich, and
  Liu}}]{Xu2016}
\bibinfo{author}{\bibfnamefont{Z.-F.} \bibnamefont{Xu}},
  \bibinfo{author}{\bibfnamefont{L.}~\bibnamefont{You}},
  \bibinfo{author}{\bibfnamefont{A.}~\bibnamefont{Hemmerich}},
  \bibnamefont{and} \bibinfo{author}{\bibfnamefont{W.~V.} \bibnamefont{Liu}},
  \bibinfo{journal}{Phys. Rev. Lett.} \textbf{\bibinfo{volume}{117}},
  \bibinfo{pages}{085301} (\bibinfo{year}{2016}),
  \urlprefix\url{http://link.aps.org/doi/10.1103/PhysRevLett.117.085301}.

\bibitem[{\citenamefont{Di~Liberto et~al.}(2016)\citenamefont{Di~Liberto,
  Hemmerich, and Morais~Smith}}]{DiLiberto2016}
\bibinfo{author}{\bibfnamefont{M.}~\bibnamefont{Di~Liberto}},
  \bibinfo{author}{\bibfnamefont{A.}~\bibnamefont{Hemmerich}},
  \bibnamefont{and}
  \bibinfo{author}{\bibfnamefont{C.}~\bibnamefont{Morais~Smith}},
  \bibinfo{journal}{Phys. Rev. Lett.} \textbf{\bibinfo{volume}{117}},
  \bibinfo{pages}{163001} (\bibinfo{year}{2016}),
  \urlprefix\url{http://link.aps.org/doi/10.1103/PhysRevLett.117.163001}.

\bibitem[{\citenamefont{Solnyshkov et~al.}(2016)\citenamefont{Solnyshkov,
  Nalitov, and Malpuech}}]{Solnyshkov2016}
\bibinfo{author}{\bibfnamefont{D.~D.} \bibnamefont{Solnyshkov}},
  \bibinfo{author}{\bibfnamefont{A.~V.} \bibnamefont{Nalitov}},
  \bibnamefont{and} \bibinfo{author}{\bibfnamefont{G.}~\bibnamefont{Malpuech}},
  \bibinfo{journal}{Phys. Rev. Lett.} \textbf{\bibinfo{volume}{116}},
  \bibinfo{pages}{046402} (\bibinfo{year}{2016}),
  \urlprefix\url{http://link.aps.org/doi/10.1103/PhysRevLett.116.046402}.

\bibitem[{\citenamefont{Solnyshkov et~al.}(2017)\citenamefont{Solnyshkov, Bleu,
  Teklu, and Malpuech}}]{Solnyshkov2017}
\bibinfo{author}{\bibfnamefont{D.~D.} \bibnamefont{Solnyshkov}},
  \bibinfo{author}{\bibfnamefont{O.}~\bibnamefont{Bleu}},
  \bibinfo{author}{\bibfnamefont{B.}~\bibnamefont{Teklu}}, \bibnamefont{and}
  \bibinfo{author}{\bibfnamefont{G.}~\bibnamefont{Malpuech}},
  \bibinfo{journal}{Phys. Rev. Lett.} \textbf{\bibinfo{volume}{118}},
  \bibinfo{pages}{023901} (\bibinfo{year}{2017}),
  \urlprefix\url{http://link.aps.org/doi/10.1103/PhysRevLett.118.023901}.

\end{thebibliography}

\appendix
\begin{widetext}

\section{Effective 2 by 2 Hamiltonian with trigonal warping}

Between the two limits considered in the text to compute analytically valley topological charges, it is still possible to write an effective 2 by 2 Hamiltonian:
\begin{equation}
H_{K,eff}=\begin{pmatrix}
\Delta_{AB} &&\frac{3a\delta J}{2}(q_y-iq_x) \\ \frac{3a\delta J}{2}(q_y+iq_x) && -\Delta_{AB}
\end{pmatrix}+ \frac{1}{\Delta_{AB}^2+9\delta J^2}\begin{pmatrix}
\frac{9a^2J^2\Delta_{AB}}{4}q^2 &&\frac{27a^2J^2\delta J}{4}(q_y+iq_x)^2 \\ \frac{27a^2J^2\delta J}{4}(q_y-iq_x)^2&& -\frac{9a^2J^2\Delta_{AB}}{4}q^2 
\end{pmatrix}
\end{equation}

\begin{figure}[H]
 \begin{center}
 \includegraphics[scale=0.49]{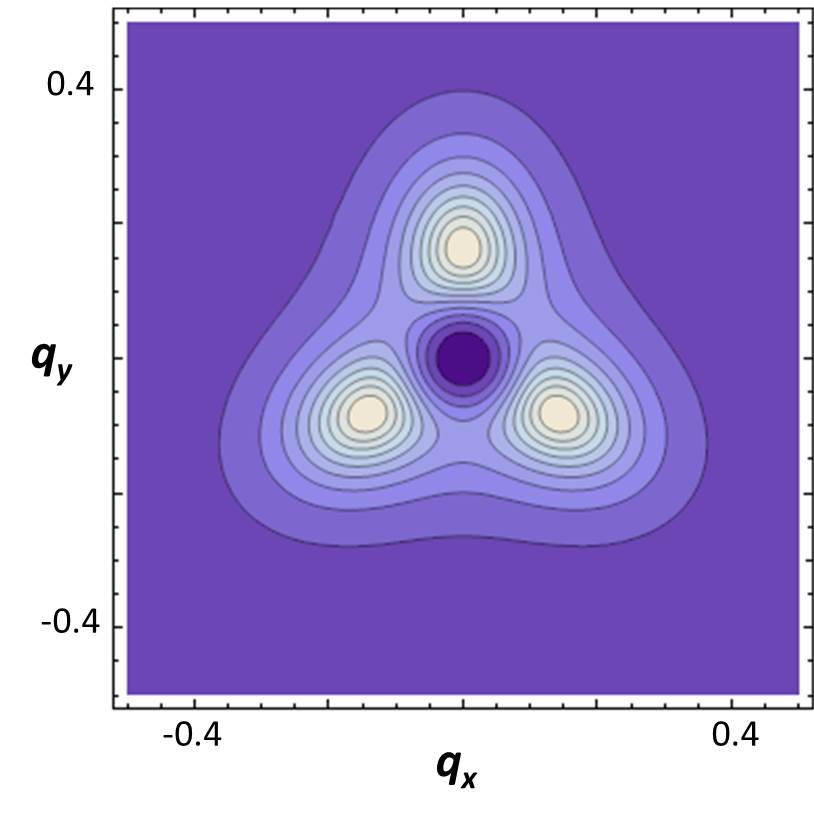}
\caption{(Color online) Berry curvature around K point with trigonal warping contribution (eq. A2). (Parameters: $\Delta_{AB}(r)=0.1 J$, $\delta J=0.2 J$)} 
  \end{center}
 \end{figure}

This effective Hamiltonian allows to compute the energy dispersion taking into account the trigonal warping. Moreover, one can also compute the Berry curvature of the low energy bands which can be written as follow in polar coordinates ($q$,$\phi$).

\begin{equation}
\mathbf{B}=-\frac{\tau_z 18 a^2 \Delta_{AB} \delta J^4 (4\delta J^4 - a^2 J^2 (\delta J^2 + 4 J^2) q^2 + 
     a^3 J^4 q^3 \mathrm{sin}(3 \phi)}{(\Delta_{AB}^2 (4 \delta J^2 + a^2 J^2 q^2)^2 + 
    9 a^2 \delta J^2 q^2 (4 \delta J^4 + a^2 J^4 q^2) + 
    36 a^3 \delta J^4 J^2 q^3 \mathrm{sin}(3 \phi))^{3/2}}
 \end{equation}

 The form of the the resulting Berry curvature describes well the additional bands extrema contributions (see Fig. (5)). This quantity is not integrable analytically but converge numerically to $2\pi$ when $\delta J>>\Delta_{AB}$. Each additional Dirac points, located at  $\vec{t}_1=\left(0,\frac{2\delta J^2}{a J^2}\right)$ , $\vec{t}_2=\left(\frac{\sqrt{3}\delta J^2}{a J^2},-\frac{\delta J^2}{a J^2}\right)$, $\vec{t}_3=\left(-\frac{\delta J^2}{a J^2},-\frac{\sqrt{3}\delta J^2}{a J^2}\right)$, carrying $\pi$ whereas the central one carries $-\pi$. However, this low energy effective theory doesn't carry information about the second valley. When $\delta J$ increases, the additional Dirac points of the $K$ and $K'$ carrying opposite topological charges move in opposite directions on the same line $KK'$ and cancel each others for some critical parameters due to the finite size of the Brillouin zone. Then, the second limit, with $C_{K,K'}^{(2)}=- \tau_z\frac{1}{2}\mathrm{sign}(\Delta_{AB})$,  discussed in the main text is achieved when the concept of valley is again well defined: that is when $\delta J$ becomes sufficiently large.
 
\end{widetext}

\end{document}